\documentclass[aps,amsmath,amssymb,showpacs,showkeys]{revtex4-2}
\usepackage[dvips]{graphicx,color}
\usepackage{times}
\usepackage{braket}
\usepackage{xcolor, soul}
\usepackage[
  colorlinks=true,
  urlcolor=blue,
  linkcolor=red,
  citecolor=blue
]{hyperref}
\usepackage{graphicx}
\usepackage{amsmath}
\usepackage{hyperref}
\hypersetup{colorlinks,citecolor=blue}
\hypersetup{colorlinks=true, linkcolor=red, filecolor=magenta, urlcolor=blue}
\usepackage{amsmath}
\usepackage{xcolor}
\usepackage{orcidlink}
\begin{document}
\title{Observational constraints on cosmological parameters in the Bianchi 
type III Universe with $\boldsymbol{f(R,T)}$ gravity theory}

\author{Pranjal Sarmah \orcidlink{0000-0002-0008-7228}}
\email[E-mail:]{p.sarmah97@gmail.com}

\author{Umananda Dev Goswami \orcidlink{0000-0003-0012-7549}}
\email[E-mail:]{umananda@dibru.ac.in}

\affiliation{Department of Physics, Dibrugarh University, Dibrugarh 786004, 
Assam, India}

\begin{abstract}
Bianchi type III (BIII) metric is an interesting anisotropic model for 
studying cosmic anisotropy as it has an additional exponential term multiplied 
to a directional scale factor. Thus, the cosmological parameters obtained 
for this BIII metric with the conventional energy-momentum tensor within the
framework of a modified gravity theory and the estimation of their values with 
the help of Hubble, Pantheon plus and other observational data may provide 
some new information in cosmic evolution. In this work, we have studied the 
BIII metric under the framework of $f(R,T)$ gravity theory and estimated the 
values of the cosmological parameters for three different models of this 
gravity theory by using the Bayesian technique. In our study, we found that 
all the models show consistent results with the current observations but show 
deviations in the early stage of the Universe. In one model we have found a 
sharp discontinuity in the {radiation}-dominated phase of the Universe. Hence 
through this study, we have found that some of the $f(R,T)$ gravity models may 
not be suitable for studying evolutions and early stages of the Universe in 
the BIII metric even though they show consistent results with the current 
observations. 
\end{abstract}

\maketitle
\section{Introduction}\label{1}
The two fundamental cosmological principles of homogeneity and isotropy 
constitute the foundation of standard cosmology, commonly referred to as 
$\Lambda$CDM cosmology. With the Friedmann-Lema\^itre-Robertson-Walker (FLRW) 
metric, backed by an energy-momentum tensor in conventional perfect fluid 
form, this theory offers solutions to numerous inquiries concerning our 
comprehension of the Universe \cite{Pebbles_1994}. Expanding the conventional 
formalism to find alternative theories as well as modifications to general 
relativity (GR) have been undertaken by researchers due to the motivation 
from various factors such as the accelerated expansion of the Universe 
\cite{Riess, Perlmutter_1999, Ma_2011}, non-observational evidence on dark 
matter (DM) \cite{Trimble_87} and dark energy (DE) \cite{Frieman_2008}, etc. 
Thus, modified theories of gravity (MTGs) are a class of suitable formalisms 
to counter the concept of exotic matter and energy content of the Universe. 

One of the simplest MTGs is the $f(R)$ theory \cite{Felice_2010,Faraoni_2010,
 Harko_2018} of gravity in which the 
Ricci scalar $R$ of Einstein Hilbert action is replaced by a function of the 
Ricci scalar. The theory is useful for studying various branches of theoretical 
research including cosmology \cite{Gogoi_2021, Li_2007,Duniya_2023, Bajardi_2022,Goswami_2014}, black hole physics \cite{Maroto_2009,Nashed_2022, Hazarika_2024,Chaturvedi_2023, Karmakar_2024}, cosmic ray 
physics \cite{SPS_20241, SPS_20242}, etc. 
Another popular MTG in cosmology is the $f(R,T)$ gravity 
\cite{Harko_2011, Harko_2018, Barrientos_2018} in which the 
gravitational Lagrangian is the arbitrary function of Ricci scalar $R$ and 
the trace of the energy-momentum tensor $T$. Thus $f(R,T)$ theory is based on 
a source term that reflects the change of the energy-momentum tensor for the 
metric. This source term has a generic formulation based on the matter 
Lagrangian $L_m$. Each $L_m$ option generates a unique set of field 
equations \cite{Kavya_2022, Jaybhaye_2024, Jaybhaye_2022}. Extensive research
has been carried out with the $f(R,T)$ theory of gravity and some of them are 
found in Refs.~\cite{Harko_2011, Jeakel_2024, Baffou_2021, Tretyakov_2018, 
Malik_2024, Rudra_2021}. Besides these MTGs, several other alternative gravity 
theories like teleparallel theories including $f(T)$ \cite{Cai_2016, 
Sabiee_2022, Dimakis_2024, Fayaz_2014, Briffa_2023}, $f(Q)$ \cite{Sarmah_2023, 
Sarmah_2024, Khyllep_2023, espo, Koussour_2023, Shabani_2024, Shahoo_2024} 
theories and some standard model extensions (SMEs) like bumblebee gravity 
theory \cite{Capelo_2015, Maluf_2021, Sarmah_2024b} have been popular among the 
researchers in cosmological studies in recent times. These MTGs and 
alternative forms of gravity theories have shown promising results when 
compared with observations \cite{Kavya_2022, Rudra_2021, Briffa_2023, 
Sarmah_2024b, Shahoo_2024}.

Besides the lack of direct observational evidence of the existence of exotic 
matter and energy content, there are also  some inadequacies while
considering standard assumptions like isotropy and homogeneity in cosmological
study. Although with these assumptions researchers have successfully explained 
most of the cosmological aspects with the help of the standard $\Lambda$CDM 
model which includes the Hubble tension \cite{Planck_2018, Nedelco_2021}, the 
$\sigma_8$ tension \cite{Planck_2018}, the coincidence problem 
\cite{Velten_2014}, etc., however, several dependable observational data 
sources, including WMAP \cite{wmap,wmap1,wmap2}, SDSS (BAO) 
\cite{SDSS_2005,Bessett_2009,Tully_2023}, and Planck 
\cite{Planck_2015,Planck_2018} have shown some deviations from
the principles of standard cosmology and hence suggest that the Universe may
have some anisotropies. Further, research suggests that the Universe has a 
large-scale planar symmetric geometry. The eccentricity of order $10^{-2}$ can 
match the quadrupole amplitude with observational evidence without altering 
the higher-order multipole of temperature anisotropy in the CMB angular power 
spectrum \cite{Tedesco_2006}. Polarization study of electromagnetic radiation 
traversing long distances confirms the presence of asymmetry axes in the 
Universe \cite{akarsu_2010}. Thus, isotropy and homogeneity assumptions alone 
cannot fully explain all cosmological aspects.

To explain anisotropies, we need a metric with a homogenous background but 
with an anisotropic feature. Luigi Bianchi proposed a class of such 
anisotropic metrics which has been classified among eleven types out of which
Bianchi type I, type III, type V and type IX are generally chosen by 
researchers to extract information on anisotropy in cosmological studies. 
However, only a very limited work has been carried out while studying the 
cosmological parameters in these metrics and most of them are restricted to 
type I. Some work of anisotropic cosmological studies in Bianchi type I metric 
have been found in Refs.~\cite{Sarmah_2022,Cea_2022, Perivolaropoulos_2014,
Berera_2004, Campanelli_2006, Campanelli_2007, Paul_2008, Barrow_1997, 
akarsu_2019}. Therefore it would be interesting to study the cosmological 
parameters by using observational data like Hubble data, BAO data, and 
Supernovae Type Ia data to understand the role of anisotropy in cosmic 
evolution in Bianchi type III (BIII) metric.
 
Various aspects of cosmological studies have been carried out using the 
BIII metric by different researchers. An accurate specific solution to the 
Einstein field equations for vacuum including a cosmological constant in 
BIII  metric has been found in Ref.~\cite{Mouss_1981}. Lorenz had proposed a 
model that includes dust and a cosmological constant in the BIII metric 
\cite{Lorenz_1982}. Another work that proposed a viscous cosmological model 
with a changeable gravitational constant ($G$) and $\Lambda$ is found in 
Ref.~\cite{chak_2001}. A work with a variable $G$ and $\Lambda$ in the presence 
of a perfect fluid, assuming a conservation rule for the energy-momentum 
tensor in the BIII metric has been studied in Ref.~\cite{Singh_2007}. The BIII 
model with a perfect fluid, time-dependent $\Lambda$, and constant deceleration 
parameter has been found in Ref.~\cite{tiwari_2009}. Letelier investigated 
certain two-fluid cosmological models with comparable symmetries to the 
BIII model, in which the separate four-velocity vectors of the two 
non-interacting perfect fluids yield an axially symmetric anisotropic 
pressure \cite{Lete_1980}. Thus the study of cosmological parameters and 
constraining their values with the help of available observational data may 
provide new insights into modern cosmology.

Here we use the BIII metric in $f(R,T)$ gravity theory to analyze the 
Universe's anisotropy and cosmological parameters. We used accessible 
observational data, including Hubble data, Pantheon Plus data, and BAO data, 
to gain a more realistic and physical understanding of the Universe. We have 
used powerful Bayesian inference techniques to estimate the cosmological 
parameters for three different $f(R,T)$ models. Based on the estimated 
values of the parameters we have further studied the effective equation of 
state $\omega_{eff}$ and the deceleration parameter. Based on the outcomes 
of the results we have made some comments on the viability of some $f(R,T)$ 
models in cosmological studies. 

The current article is organized as follows. Starting the introduction part 
in Section \ref{1} to explain the importance of Bianchi III Universe and 
MTGs as well as the alternative theories of gravity, specifically the $f(R,T)$ 
theory from various literatures and then we have discussed the general form of 
field equations in $f(R,T)$ gravity theory in Section \ref{2}. In Section 
\ref{3}, we have developed the required field equations and continuity 
equation for the BIII metric and also derived the cosmological parameters for 
three different $f(R,T)$ models. In Section \ref{4}, we have constrained the 
model parameters and cosmological parameters by using the techniques of 
Bayesian inference by using various observational data compilations and 
comparing our models' results with the standard cosmology by using the 
constrained values of the parameters for three $f(R,T)$ models. Finally, the 
article has been summarised with conclusions in Section \ref{5}. 
\section{$\boldsymbol{f(R,T)}$ gravity theory and field equations}\label{2}

The modified Einstein-Hilbert action for the $f(R,T)$ theory of gravity is 
\cite{Harko_2011}
\begin{equation}\label{action}
{S} = {\frac{1}{2\kappa}} \int {f(R,T)} \sqrt{-{g}} ~ {d^{4}} x + \int {\mathcal{L_M}} \sqrt{-{g}} ~ {d^{4}x}
\end{equation}
where $\kappa = 8\pi G$ and  $\mathcal{L_M}$ is the matter Lagrangian density. 
The corresponding field equations obtained through varying the action 
\eqref{action} with respect to metric tensor $g_{\mu \nu}$ can be written as
\begin{align}\label{FE}
f_{R}(R,T)R_{\mu \nu} - \frac{1}{2}f(R,T)g_{\mu \nu} + (g_{\mu \nu} \square- \nabla_{\mu}\nabla_{\nu})f_{R}(R,T)=\kappa T_{\mu\nu}-f_{T}(R,T)T_{\mu\nu} - f_{T}(R,T)\Theta_{\mu \nu}.
\end{align}
Here, $f_R(R,T)$ and $f_T(R,T)$ represent the derivatives of $f(R,T)$ with 
respect to $R$ and $T$ respectively. $T_{\mu\nu}$ is the energy-momentum 
tensor and is expressed as 
\begin{equation}\label{BE}
T_{\mu \nu} = -\frac{2}{\sqrt{-g}}\frac{\delta \left( \sqrt{-g}\,\mathcal{L_M}\right)}{\delta g^{\mu \nu}}.
\end{equation}
Further, the term $\Theta_{\mu \nu}$ in equation \eqref{FE} can be
written as \cite{Harko_2011}
\begin{equation}\label{theta}
\Theta_{\mu \nu} = -2T_{\mu\nu
}+ g_{\mu\nu} \mathcal{L_M}-2g^{\alpha\beta} \frac{\partial^{2}\mathcal{L_M}}{\partial g^{\mu\nu} \partial g^{\alpha \beta}}.
\end{equation}

With the help of equation \eqref{FE} we will derive field equations for the
BIII metric for the conventional perfect fluid case in the next section for 
some specific $f(R,T)$ models.

\section{Bianchi III Cosmology in $\boldsymbol{f(R,T)}$ Gravity Theory}\label{3}
We have considered the BIII metric in our study which has the form:
\begin{equation}\label{metric}
ds^2 = -\,dt^2+a_1^2(t) dx^2+ a_2^2(t)e^{-2mx} dy^2+ a_3^2(t) dz^2.
\end{equation} 
Here, $a_1$, $a_2$, and $a_3$ are functions of time and can be considered as 
the scale factors in $x$, $y$, and $z$ directions respectively. $m$ is a 
constant. Hence, this metric provides three directional Hubble parameters 
$H_1 =\dot{a_1}/{a_1}$, $H_2 =\dot{a_2}/{a_2}$, $H_3 = \dot{a_3}/{a_3}$
along three different directions. Thus, the average expansion scale factor for 
this metric is $(a_1 a_2 a_3)^{\frac{1}{3}}$ \cite{Sarmah_2022} and the average Hubble parameter can be written as
\begin{equation}\label{Hub}
H = \frac{1}{3}(H_1+H_2+H_3).
\end{equation}
For the perfect fluid matter-energy density model of the Universe, the 
energy-momentum tensor $T_{\mu}^{\nu} = diag(-\rho,P,P,P)$ and hence the 
components of ${\Theta_{\mu\nu}}$ from equation \eqref{theta} can be written as
${\Theta_{00} = -2\rho + p}$, ${ \Theta_{11} = -3a_1^{2}\,p}$, 
${ \Theta_{22} = -3a_2^{2}\,e^{-2mx}\,p}$, 
${ \Theta_{33} = -3a_3^{2}\,p}$ for ${\mathcal{L_M}=-\,p}$.

Now we are ready for deriving field equations in the BIII metric for different 
$f(R,T)$ models. In our work, we have considered the following three $f(R,T)$ 
models as suggested in Ref.~\cite{Harko_2011}:
\begin{equation}
f(R, T) = 
\begin{cases} 
f_1(R) + f_2(T),\\
R + 2f(T),  \\
f_1(R) + f_2(R)f_3(T). 
\end{cases}
\end{equation}
Each of these models is considered for our purpose as follows:
   
\subsection{$f(R,T) = f_1(R) + f_2(T)$}
The considered form of $f(R,T)$ is a standard form of the $f(R,T)$ gravity 
models. In this study, we have considered $f(R,T) = \alpha R + \beta f(T)$ for our analysis, where $\alpha$ and $\beta$ are two constants. The 
metric-independent form of the field equations for the considered model can be 
written as
\begin{equation}\label{FEn}
\alpha \Big(R_{\mu\nu} -\frac{1}{2}g_{\mu\nu}R\Big) = \big[\kappa + \beta f_{T}(T)\big]T_{\mu\nu} +\Big[\beta\, p f_{T}(T)+\frac{1}{2} \beta f(T)\Big]g_{\mu\nu} {=\kappa T^{eff}_{\mu\nu}},
\end{equation}
{where} ${T^{eff}_{\mu\nu}}$ {represents the effective 
energy-momentum tensor for the considered model, which can be expressed as}
${T^{eff}_{\mu\nu} = T_{\mu\nu} + \tilde T^{\mu\nu}}$ {with}
\begin{equation}\label{EMTt}
{\kappa \tilde T^{\mu\nu} = \beta f_{T}(T)T_{\mu\nu} +\Big[\beta\, p f_{T}(T)+\frac{1}{2} \beta f(T)\Big]g_{\mu\nu}}.
\end{equation} 
For the BIII metric with $f(T) = \lambda T$ in which $\lambda$ is a constant 
along with the considered conventional energy-momentum tensor, the set of field equations \eqref{FEn} take the form:
\begin{align}\label{fet}
{\frac{\dot{a_1}\dot{a_2}}{a_1 a_2}+\frac{\dot{a_2}\dot{a_3}}{a_2 a_3}+\frac{\dot{a_3}\dot{a_1}}{a_3 a_1}  - \left(\frac{m}{a_1}\right)^2} & {= \frac{1}{\alpha}\left(\kappa + \frac{3}{2} \lambda \beta \right)\rho - \frac{5\lambda \beta}{2 \alpha}\, p,}\\[5pt]
\label{fex}
{\frac{\ddot{a_2}}{a_2}+\frac{\ddot{a_3}}{a_3}+ \frac{\dot{a_2}\dot{a_3}}{a_2 a_3}} & = {-\, \frac{1}{\alpha}\left(\kappa + \frac{7}{2} \lambda \beta \right)p + \frac{\lambda \beta}{2 \alpha}\, \rho,}\\[5pt]
\label{fey}
 {\frac{\ddot{a_3}}{a_3}+\frac{\ddot{a_1}}{a_1}+ \frac{\dot{a_3}\dot{a_1}}{a_3 a_1} }& = {-\, \frac{1}{\alpha}\left(\kappa + \frac{7}{2} \lambda \beta \right)p + \frac{\lambda \beta}{2 \alpha}\, \rho,}\\[5pt]
\label{fez}
{\frac{\ddot{a_1}}{a_1}+\frac{\ddot{a_2}}{a_2}+ \frac{\dot{a_1}\dot{a_1}}{a_1 a_2}-\left(\frac{m}{a_1}\right)^2 }& = {-\, \frac{1}{\alpha}\left(\kappa + \frac{7}{2} \lambda \beta \right)p + \frac{\lambda \beta}{2 \alpha}\, \rho,}\\[5pt]
\label{fexy}
m\left(\frac{\dot{a_1}}{a_1}-\frac{\dot{a}_2}{a_2}\right) & = 0.
 \end{align} 
From equation \eqref{fexy}, we have observed that for $m\neq 0$, $H_1 = H_2$ 
and with this condition along with the consideration $\kappa = 1$ we can 
rewrite the above field equations as
\begin{align}\label{fe1}
{H_1^2 + 2 H_1 H_3 - \left(\frac{m}{a_1}\right)^2} & = {\frac{1}{\alpha}\left(1 + \frac{3}{2} \lambda \beta \right)\rho - \frac{5\lambda \beta}{2 \alpha}\, p,}\\[5pt]
\label{fe2}
{H_1^2+ H_3^2 + H_1 H_3 + \left(\dot{H_1} + \dot{H_3} \right) }& = {-\,\frac{1}{\alpha}\left(1 + \frac{7}{2} \lambda \beta \right)p + \frac{\lambda \beta}{2 \alpha}\, \rho,} \\[5pt]
\label{fe3}
{3H_1^2+ 2\dot{H_1} -\left(\frac{m}{a_1}\right)^2} & = {-\, \frac{1}{\alpha}\left(1 + \frac{7}{2} \lambda \beta \right)p + \frac{\lambda \beta}{2 \alpha}\, \rho.}
\end{align}
Moreover, for the condition $m=0$ equations \eqref{fet}, \eqref{fex},
\eqref{fey} and \eqref{fez} can be rewritten as
 \begin{align}
{H_1 H_2+H_2 H_3 + H_3 H_1} & = {\frac{1}{\alpha}\left(1 + \frac{3}{2} \lambda \beta \right)\rho - \frac{5\lambda \beta}{2 \alpha}\, p,}\\[5pt]
{H_2^2+ H_3^2 + H_2 H_3 + \left(\dot{H_2} + \dot{H_3} \right) }& = {-\,\frac{1}{\alpha}\left(1 + \frac{7}{2} \lambda \beta \right)p + \frac{\lambda \beta}{2 \alpha}\, \rho,}\\[5pt]
{H_1^2+ H_3^2 + H_1 H_3 + \left(\dot{H_1} + \dot{H_3} \right)} & = {-\,\frac{1}{\alpha}\left(1 + \frac{7}{2} \lambda \beta \right)p + \frac{\lambda \beta}{2 \alpha}\, \rho,}\\[5pt]
{H_1^2+ H_2^2 + H_1 H_2 + \left(\dot{H_1} + \dot{H_2} \right)} & = {-\,\frac{1}{\alpha}\left(1 + \frac{7}{2} \lambda \beta \right)p + \frac{\lambda \beta}{2 \alpha} \rho.}
\end{align}
Further, for the condition of $m = 0$ and $H_1 = H_2$, these field equations 
can be rewritten as
\begin{align}
 {H_1^2+ 2H_3 H_1 }& = {\frac{1}{\alpha}\left(1 + \frac{3}{2} \lambda \beta \right)\rho - \frac{5\lambda \beta}{2 \alpha}\, p,}\\[5pt]
 {H_1^2+ H_3^2 + H_1 H_3 + \left(\dot{H_1} + \dot{H_3} \right)} & = {-\,\frac{1}{\alpha}\left(1 + \frac{7}{2} \lambda \beta \right)p + \frac{\lambda \beta}{2 \alpha}\, \rho,}\\[5pt]
{3H_1^2 + 2\dot{H_1}} & = {-\,\frac{1}{\alpha}\left(1 + \frac{7}{2} \lambda \beta \right)p + \frac{\lambda \beta}{2 \alpha}\, \rho.}
\end{align}
 
It is to be noted that the $m =0$ and $H_1 = H_2$ conditions reduce the BIII 
Universe to LRS-BI Universe and for $m=0$, $H_1 \ne H_2$ it reduces to 
standard BI Universe. Since the motive of this work is to explore the BIII 
Universe, thus we are only interested in looking at the scenario of 
$m\neq 0$ and $H_1=H_2$ in our study. The shear scalar $\sigma^2$ for the 
considered scenario can be written as
 \begin{equation}
 \sigma^2 = \frac{1}{3}\left(H_1-H_3\right)^2.
 \end{equation}

For the conditions $m \neq 0$ and $H_1 = H_2$, we have already derived the 
field equations \eqref{fe1}, \eqref{fe2} and \eqref{fe3}. Now considering
a relation $H_3 = \gamma H_1$ through using $\theta^2 \propto \sigma^2$ 
condition \cite{Sarmah_2023, Sarmah_2024} in which $\theta^2$ and $\sigma^2$ 
are the expansion scalar and shear scalar respectively, we can rewrite these 
field equations as
 \begin{align}
{3H^2}& = {\frac{(2+\alpha)^2}{3\alpha(1+2\gamma)} \left[(1+\frac{3}{2}\lambda \beta)\rho -\frac{5}{2}\lambda \beta p + \frac{\alpha m^2}{a^{\frac{6}{2+\gamma}}}\right]}\label{Fen1},\\[8pt]
{3H^2 + \frac{2}{3}\left(2+\gamma\right)\dot{H} }& = {-\,\frac{(2+\gamma)^2}{9\alpha}\left[(1+\frac{7}{2}\lambda \beta)p -\frac{1}{2}\lambda \beta \rho - \frac{\alpha m^2}{a^{\frac{6}{2+\gamma}}}\right].}\label{Fen2}
 \end{align}

{Since in $f(R,T)$ gravity theory the conventional energy-momentum 
tensor is not conserved \cite{O_2014, Carvalho_2020, Santos_2019, Moraes_2018}, the continuity equation can be obtained by using the condition of the 
conservation of the effective energy-momentum tensor, 
i.e.~}\\
${\nabla_{\mu}T^{\mu\nu}_{eff}= \nabla_{\mu}T^{\mu\nu}+\nabla_{\mu}{\tilde T^{\mu\nu}} = 0}$ \cite{Moraes_2018} {as}
\begin{equation}\label{cont}
{
\dot{{\rho}} = -\,\frac{3{H}\rho\left\{1-{{\lambda} {\beta}}\right\}(1+{\omega})}{\left\{1+\frac{{\lambda} {\beta}}{2}(5{\omega}-3)\right\}}.}
\end{equation}
{Here, the components} 
${\tilde{T}^{0}_{0} =\frac{\lambda \beta}{2}(-3\rho+5p)}$, 
${\tilde{T}^{1}_{1} =\frac{\lambda \beta}{2}(\rho-7p)}$, 
${\tilde{T}^{2}_{2} =\frac{\lambda \beta}{2}(\rho-7p)}$ {and}
${\tilde{T}^{3}_{3} =\frac{\lambda \beta}{2}(\rho-7p)}$ 
{are obtained from relation \eqref{EMTt}.}

Using equations \eqref{Fen1} and \eqref{Fen2} we can now obtain the effective 
equation of state and it takes the form:
\begin{equation}\label{Om_eff}
{
\omega_{eff} = -\left( 1+\frac{2(2+\gamma)}{9}\frac{\dot{H}}{H^2}\right)=\frac{(2+\gamma)^2(1+2\gamma)}{3(2+\alpha)^2}\left[\frac{(1+\frac{7}{2}\lambda \beta)p -\frac{1}{2}\lambda \beta \rho - \alpha\, m^2 a^{-\frac{6}{2+\gamma}}}{(1+\frac{3}{2}\lambda \beta)\rho -\frac{5}{2}\lambda \beta p + \alpha\, m^2a^{-\frac{6}{2+\gamma}}}\right].}
\end{equation}
{This effective equation state is the ratio of the effective pressure 
to the effective density. The term appearing on right hand side of equation 
\eqref{Fen1} can be considered to be the effective density if we compare it 
with the standard Friedmann equation and similarly the term appearing on the 
right hand side of equation \eqref{Fen2} can be considered to be the effective 
pressure in our study.}

Further, the deceleration parameter can be written as
 \begin{equation}\label{dec}
 q =-\left(1+\frac{\dot{H}}{H^2}\right) = \frac{(5-2\gamma)+9\,\omega_{eff}}{2(2+\gamma)}.
 \end{equation}
Applying the condition $p = \omega \rho$ in which $\omega$ is the equation of 
state with $\omega = 0$ for matter, $\omega = \frac{1}{3}$ for radiation and 
$\omega = -1$ for dark energy, the solution for $\rho$ from the continuity 
equation \eqref{cont} can be written as
 \begin{equation}\label{rho}
{ \rho = \rho_0 (1+z)^{\frac{3(1+\omega)\{1-{\lambda \beta}\}}{\{1+\frac{\lambda \beta}{2}(5\omega-3)\}}}}.
 \end{equation} 
Here we have taken $a = \frac{1}{(1+z)}$ in which $z$ is the cosmological 
redshift. Thus, the  Hubble parameter from equiation \eqref{Fen1} can be 
expressed as
\begin{align}\label{hub_new1}
{
H}&{(z) = H_0\sqrt{\frac{(2+\alpha)^2}{3\alpha(1+2\gamma)}}\,\times}
\nonumber\\[5pt]
& {\sqrt{\left[(1+\frac{3}{2}\,\lambda \beta)\,\Omega_{m0}\,(1+z)^{\frac{3\{1-{\lambda \beta}\}}{\{1-\frac{3}{2}\lambda \beta\}}} +(1+\frac{2}{3}\,\lambda \beta)\,\Omega_{r0}\,(1+z)^{\frac{4\{1-{\lambda \beta}\}}{\{1-\frac{2}{3}\lambda \beta\}}}+(1+4\lambda \beta)\,\Omega_{\Lambda 0}+ {\alpha\, m^2}{(1+z)^{\frac{6}{2+\gamma}}}\right]},}
\end{align}
where $\Omega_{m0} = {\rho_{m0}}/{3H_{0}^2}$, 
$\Omega_{r0} = {\rho_{r0}}/{3H_{0}^2}$ and 
$\Omega_{\Lambda0} = {\rho_{\Lambda0}}/{3H_{0}^2}$ are the density parameters 
for matter, radiation and dark energy respectively. Further, with the help of 
equation \eqref{hub_new1}, we can derive the distance modulus by using the 
equation,
 \begin{equation}\label{Dm}
 D_m = 5\log d_L+ 25,
 \end{equation}
in which $d_L$ is the luminosity distance and it can be derived by using the 
expression, 
\begin{equation}\label{d_L}
 d_L = (1+z) \int_{0}^{\infty}\!\! \frac{dz}{H(z)}.
 \end{equation}
Moreover, the equation \eqref{Om_eff} can be rewritten as
 \begin{align}\label{omf_new1}
{\omega_{eff}}&{(z) = \frac{(2+\gamma)^2(1+2\gamma)}{3(2+\alpha)^2}\,\times}\nonumber\\[5pt]
&{\left[\frac{-\frac{1}{2}\lambda \beta\,\Omega_{m0}(1+z)^{\frac{3\{1-{\lambda \beta}\}}{\{1-\frac{3}{2}\lambda \beta\}}}+\frac{1}{3}(1+{2}\lambda \beta)\,\Omega_{r0}(1+z)^{\frac{4\{1-{\lambda \beta}\}}{\{1-\frac{2}{3}\lambda \beta\}}}-(1+4\lambda\beta)\Omega_{\Lambda 0} -{\alpha\, m^2}{(1+z)^{\frac{6}{2+\gamma}}}}{(1+\frac{3}{2}\,\lambda \beta)\,\Omega_{m0}\,(1+z)^{\frac{3\{1-{\lambda \beta}\}}{\{1-\frac{3}{2}\lambda \beta\}}} +(1+\frac{2}{3}\,\lambda \beta)\,\Omega_{r0}\,(1+z)^{\frac{4\{1-{\lambda \beta}\}}{\{1-\frac{2}{3}\lambda \beta\}}}+(1+4\lambda \beta)\,\Omega_{\Lambda 0}+ {\alpha\, m^2}{(1+z)^{\frac{6}{2+\gamma}}}}\right].}
\end{align}
 
With these, the required set of equations and cosmological parameters are ready 
for further analysis. One of the major tasks from here is to constrain the 
different cosmological and model parameters that appear in different 
cosmological expressions. The detailed methods of parameter constraining have 
been discussed in later sections. 
\subsection{$f(R,T) = R + 2 f(T)$}
This model is the simplified version of the previous 
$f(R,T) = \alpha R + \beta f(T)$ model with $\alpha =1$, $\beta = 2$ and 
$f(T) = \lambda T$. Thus the field equations for this considered model can be 
written for $m\neq 0$ and $H_1 = H_2$ and $H_3 = \gamma H_1$ as
\begin{eqnarray}
{3H^2 = \frac{3}{(1+2\gamma)} \left[(1+3\lambda)\rho - 5\lambda p + \frac{ m^2}{a^{\frac{6}{2+\gamma}}}\right]}\label{hub},\\[8pt]
{3H^2 + \frac{2}{3}\left(2+\gamma\right)\dot{H}=-\frac{(2+\gamma)^2}{9}\left[(1+7\lambda)p - \lambda \rho - \frac{ m^2}{a^{\frac{6}{2+\gamma}}}\label{Fe2}\right].}
\end{eqnarray}
{Here, equations \eqref{cont} and \eqref{rho} are reduced to the 
following forms as}
\begin{align}\label{cont_2}
{\dot{\rho}} & {= -\,\frac{3H\rho(1+\omega)\{1-{2\lambda}\}}{\{1+{\lambda}(5\omega-3)\}}},\\[8pt]
\label{rho_2}
{\rho} &{= \rho_0 (1+z)^{\frac{3(1+\omega)\{1-{2\lambda}\}}{\{1+{\lambda}(5\omega-3)\}}}.}
\end{align} 
The Hubble parameter for the considered model takes the form:
\begin{equation}\label{hub_new2}
H(z) = H_0\sqrt{\frac{3}{(1+2\gamma)}\left[(1+3\lambda)\,\Omega_{m0}(1+z)^{\frac{3(1-2\lambda)}{(1-3\lambda)}} +(1+\frac{4}{3}\lambda)\,\Omega_{r0}(1+z)^{\frac{4(1-2\lambda)}{(1-\frac{4}{3}\lambda)}} +(1+8\lambda)\,\Omega_{\Lambda 0}+ { m^2}{(1+z)^{\frac{6}{2+\gamma}}}\right]}. 
\end{equation}
Similarly, the expression of the effective equation of state given in equation 
\eqref{omf_new1} now reduced to 
  \begin{align}\label{omf_new}
{\omega_{eff}}&{(z) = \frac{(2+\gamma)^2(1+2\gamma)}{27}\,\times}\nonumber\\[5pt]
&{\left[\frac{-\lambda\,\Omega_{m0}(1+z)^{\frac{3(1-2\lambda)}{(1-3\lambda)}}+\frac{1}{3}(1+4\lambda)\,\Omega_{r0}(1+z)^{\frac{4
(1-2\lambda)}{(1-\frac{4}{3}\lambda)}}-(1+8\lambda)\,\Omega_{\Lambda 0} -{ m^2}{(1+z)^{\frac{6}{2+\gamma}}}}{(1+3\lambda)\,\Omega_{m0}(1+z)^{\frac{3
(1-2\lambda)}{(1-3\lambda)}} +(1+\frac{4}{3}\lambda)\,\Omega_{r0}(1+z)^{\frac{4
(1-2\lambda)}{(1-\frac{4}{3}\lambda)}}  +(1+8\lambda)\,\Omega_{\Lambda 0}+ { m^2}{(1+z)^{\frac{6}{2+\gamma}}}}\right].}
 \end{align}
Further, the expressions of the deceleration parameter,  distance modulus and 
luminosity distance can be obtained for this model using equations \eqref{dec}, 
\eqref{Dm} and \eqref{d_L} respectively.
  
\subsection{$f(R,T)= f_1(R)+ f_2(R)f_3(T)$}
In this model we have considered $f_1(R) = \zeta R$, $f_2(R) = \tau R$ and 
$f_3(T) = \eta\,T$, where $\zeta$, $\tau$ and $\eta$ are some other constants.  
Thus the considered $f(R,T))$ model takes the form: $f(R,T) = \zeta R + 
\eta\, \tau R\, T = (\zeta + \eta\, \tau\, T)R$. For this form of the model 
the metric independent field equations \eqref{FE} become,
\begin{equation}\label{FE2}
  \left(\zeta+\eta\, \tau\, T\right)R_{\mu \nu} - \frac{1}{2} \left(\zeta+\eta\, \tau\, T\right)R g_{\mu\nu} = \kappa T_{\mu \nu} - \eta\, \tau R T_{\mu \nu} - \eta\, \tau R \Theta_{\mu \nu}.
  \end{equation}
Thus for the considered BIII metric, field equations \eqref{FE2} in geometric 
unit under the condition $m \neq 0$ and $H_1 = H_2$ can be written in temporal 
and spatial components as
\begin{align}\label{Fe1}
{H_1^2 + 2 H_1 H_3 - \left(\frac{m}{a_1}\right)^2} & {=\frac{1}{ \left(\zeta+\eta \tau T\right)}\left[\rho + ~\eta \tau R (\rho - p)\right],}\\[5pt]
\label{Fe2}
{H_1^2+ H_3^2 + H_1 H_3 + \left(\dot{H_1} + \dot{H_3} \right)} & {= -\, \frac{1}{ \left(\zeta+\eta \tau T\right)}\left[1 + 2 ~\eta \tau R \right]p,} \\[5pt]
\label{Fe3}
{3H_1^2+ 2\dot{H_1} -\left(\frac{m}{a_1}\right)^2 }& {= -\, \frac{1}{ \left(\zeta+\eta \tau T\right)}\left[1 + 2 ~\eta \tau R  \right]p.}
\end{align}
Like in the case of the previous model, we have considered the 
$\sigma^2 \propto \theta^2$ assumption for which here we have taken 
$H_3 = \gamma H_1$. Apart from that, with the consideration of equations 
\eqref{cont} and \eqref{rho}, and standard definitions of various density 
parameters as mentioned in the previous model, as well as considering the 
relation $p = \omega \rho$, we can derive the cosmological parameters for this 
model too. However, before deriving the cosmological parameters we have to 
rewrite the field equations in a more convenient form as follows:
\begin{align}\label{Fei}
{3H^2} & {= \frac{(2+\gamma)^2}{3(1+2\gamma)}\left[\frac{\rho + ~\eta \tau R (\rho - p)}{\zeta +\eta\, \tau T} +\frac{m^2}{a^{\frac{6}{2+\gamma}}}\right],}\\[5pt]
{3H^2+ \frac{2}{3}(2+\gamma) \dot{H} } & {= -\,\frac{(2+\gamma)^2}{9}\left[\frac{(1+2\eta \tau R)p}{\left(\zeta+\eta\, \tau T\right)}+\frac{m^2}{a^{\frac{6}{2+\gamma}}}\right].}\label{Feii}
\end{align}

{As mentioned already, here also the effective energy-momentum tensor 
conservation condition, ${\nabla_{\mu}T^{\mu\nu}_{eff} = 
\nabla_{\mu}T^{\mu\nu}+}$ ${\nabla_{\mu}{\tilde T^{\mu\nu}} = 0}$ 
is used to obtain the continuity equation. For this model, 
${T^{00}_{eff} = \frac{\rho+\eta \tau R (\rho - p)}{ \left(\zeta+\eta \tau T\right)}}$, ${ T^{ii}_{eff} = -\,g^{ii}\frac{(1 + 2\,\eta \tau R)p}{ \left(\zeta+\eta \tau T\right)}}$, where $i=(1,2,3)$. 
Thus the continuity equation for this model takes the form:}
\begin{align}\label{cont_3}
{\frac{3 \dot{a}}{a}}& {\left[{(\omega +1)\rho-\frac{2 \eta  \tau \rho^2 (1-3 \omega)^2}{\zeta +2 \eta  \tau  (5 \omega -1)\rho}}\right]+ \frac{2 \eta ^2 \tau ^2 (1-\omega) (1 -3 \omega) \left(5 \omega-1\right)\dot{\rho}\rho^2}{\{\zeta +2 \eta  \tau  (5 \omega -1) \rho\}^2}}\nonumber\\[5pt]
&{-\frac{2\eta  \tau  (1-3 \omega )\left(1-\omega\right)\rho \dot{\rho}}{\zeta +2 \eta  \tau  (5 \omega -1)\rho}+\dot{\rho} - \frac{\eta  \tau  \left(3 \omega -1\right)\dot{\rho} \left(\rho-\frac{\eta  \tau  (1 -3 \omega) (1-\omega)\rho^2}{\zeta +2 \eta  \tau  (5 \omega -1)\rho}\right)}{\{\zeta +\eta  \tau  (3 \omega -1)\rho\}}=0}.
\end{align}
{As the form of this equation \eqref{cont_3} is quite difficult to 
solve, we have reduced this equation to three different equations for 
$\omega = 0$, $\omega = 1/3$ and $\omega = -1$ respectively as follows:}
\begin{align}\label{omega0}
{\frac{3\dot{a}}{a} \left(\rho_m-\frac{\eta  \tau  \rho_m^2}{\zeta -2 \eta  \tau  \rho_m}\right)-\dot{\rho_m}\left[1+\frac{2 \eta ^2 \tau ^2 \rho_m^2}{(\zeta -2 \eta  \tau  \rho_m)^2}-\frac{2 \eta  \tau  \rho_m}{\zeta -2 \eta  \tau  \rho_m}+\frac{\eta  \tau  \left(\rho_m-\frac{\eta  \tau  \rho_m^2}{\zeta -2 \eta  \tau  \rho_m}\right)}{(\zeta -\eta  \tau  \rho_m)}\right]=0,}&\\[8pt]
\label{omegar}
{4\,\frac{\dot{a}}{a}\,\rho_r+{\dot{\rho_r}}=0,}&\\[8pt]
\label{omegal}
{\frac{3 \dot{a}}{a} \left(-\frac{16 \eta  \tau  \rho_\Lambda^2}{\zeta -12 \eta  \tau  \rho_\Lambda}\right)-\dot{\rho_\Lambda}\left[1+\frac{96 \eta ^2 \tau ^2 \rho_\Lambda^2 }{(\zeta -12 \eta  \tau  \rho_\Lambda)^2}-\frac{16 \eta  \tau  \rho_\Lambda }{\zeta -12 \eta  \tau  \rho_\Lambda}+1+\frac{4 \eta  \tau   \left(\rho_\Lambda-\frac{8 \eta  \tau  \rho_\Lambda^2}{\zeta -12 \eta  \tau  \rho_\Lambda}\right)}{(\zeta -4 \eta  \tau  \rho_\Lambda)}\right]=0}.&
\end{align}
{Here, $\rho_m$, $\rho_r$ and $\rho_\Lambda$ represent the densities 
for matter, radiation and dark energy phases of the Universe respectively 
corresponding to $\omega = 0$, $\omega = \frac{1}{3}$ and $\omega = -1$. The 
solutions of equations \eqref{omega0}, \eqref{omegar} and \eqref{omegal} are 
as given below:}
\begin{align}\label{solm}
{\rho_m}& {= \frac{\zeta  a^3+3 \zeta  \eta  \tau  e^{c_1 \zeta }\pm \zeta  \sqrt{-6 \eta  \tau  a^3 e^{c_1 \zeta }+a^6+\eta ^2 \tau ^2 e^{2 c_1 \zeta }}}{2 \left(3 \eta  \tau  a^3+2 \eta ^2 \tau ^2 e^{c_1 \zeta }\right)}},\\[8pt]
\label{solr}
{\rho_r}& {= \frac{c_1}{a(t)^4} = \rho_{r0} a^{-4}},\\[8pt]
\label{soll}
{\rho_\Lambda}& {= \log(\text{constant~term}) = \rho_{\Lambda0}.}
\end{align}
{Equation \eqref{solm} can further be reduced by neglecting the higher 
order term of $\eta$ and $\tau$ as we are expecting small deviations of the 
model's result from the GR result and thus we can write for the negative sign 
before the square root term as}
\begin{equation}\label{solmn}
{\rho_m = \zeta \rho_{m0}\, a^{-3}}.
\end{equation}
{The solution with the positive sign before the square root term gives 
a constant value and that is why that solution is not considered in this study.}
Now using equation \eqref{Fei} we can write the Hubble parameter as
\begin{align}\label{hub_new3}
{H(z)} & {= H_0\sqrt{\frac{(2+\gamma)^2}{3(1+2\gamma)}}}
\,\times\nonumber\\[5pt] 
&{\sqrt{\left[\frac{(1+ \eta \tau R)\,\zeta \Omega_{m0}(1+z)^3 + (1+\frac{2}{3}\,\eta\tau R)\,\Omega_{r0}(1+z)^4+(1+2\,\eta \tau R)\,\Omega_{\Lambda0}}
 {\zeta-3\,\eta\, \tau H_0^2 (\zeta \Omega_{m0}(1+z)^3+ 4\,\Omega_{\Lambda0})}\right]+\frac{m^2 (1+z)^{\frac{6}{2+\gamma}}}{3H_0^{2}}}.}
 \end{align}
This expression of the Hubble parameter is Ricci scalar $R$ dependent, which 
can be written in terms of density parameters. The expression of $R$ for the 
considered $f(R,T)$ model in the BIII Universe can be written as 
{(the derivation is presented in the appendix section)}
 \begin{equation}\label{R}
{R(z) = \frac{3H_0^2\left\{\zeta \Omega_{m0}(1+z)^3 + 4\,\Omega_{\Lambda0}\right\}}{\zeta + 6\eta\, \tau H_0^2\left\{-\zeta \Omega_{m0}(1+z)^3 +\frac{2}{3}\,\Omega_{r0}(1+z)^4 -6\,\Omega_{\Lambda 0}\right\}}.}
\end{equation}   
Further, the effective equation of state for the field equations \eqref{Fei} 
and \eqref{Feii} can be written as
\begin{align}\nonumber
{\omega_{eff}(z) }& {= -\left(1+ \frac{2}{9}\,(2+\gamma)\frac{\dot{H}}{H^2}\right)=\frac{(1+2\gamma)}{3}\left[\frac{(1+2\,\eta \tau R)p+ {m^2\left(\zeta + \eta\, \tau T\right)}{a^{-\frac{6}{(2+\gamma)}}}}{\left\{1+\eta\, \tau R)(1-\omega)\right\}\rho+ {m^2\left(\zeta + \eta\, \tau T\right)}{a^{-\frac{6}{(2+\gamma)}}}}\right]}\\[10pt]
& {=\frac{(1+2\gamma)}{3}\left[\frac{3H_0^2(1+2\,\eta \tau R)(\frac{1}{3}\,\Omega_{r0}(1+z)^4-\Omega_{\Lambda0})+ {m^2\left(\zeta + \eta\, \tau T\right)}{(1+z)^{\frac{6}{(2+\gamma)}}}}{(1+ \eta \tau R)\,\zeta \Omega_{m0}(1+z)^3
 + (1+\frac{2}{3}\,\eta\tau R)\,\Omega_{r0}(1+z)^4+(1+2\,\eta \tau R)\,\Omega_{\Lambda0}+ W}\right],}
\label{Om_eff3}
\end{align}
{where} 
$${W = \frac{ m^2\left(\zeta + \eta\, \tau T\right)(1+z)^{\frac{6}{(2+\gamma)}}}{3H_0^2}.}$$
It is seen that $\omega_{eff}$ also depends on $R$ and $T$. The expression of 
$R$ is already derived in equation \eqref{R} and $T$ for the considered 
energy-momentum tensor can be written as
\begin{equation}
{T = -\,\rho + 3p = -\,3H_0^2\left(\zeta \Omega_{mo}(1+z)^3 + 4\,\Omega_{\Lambda0}\right).}
\end{equation}
Similar to $\omega_{eff}$, the deceleration parameter can be derived from 
equation \eqref{dec}. Other cosmological parameters like luminosity distance 
and distance modulus can be calculated by using equations \eqref{d_L} and 
\eqref{Dm} respectively.

We are now ready to constrain the cosmological parameters and model parameters 
for graphical visualization of the parameters along with observational data. 
For this purpose, we have used a powerful Bayesian inference technique which 
we have carried out in our next section.
 \section{Parameters' estimations and constraining}\label{4}
As mentioned earlier we have employed the Bayesian inference technique for 
the estimation and constraining of cosmological parameters for all three 
$f(R,T)$ models considered here. This technique is based on Bayes theorem, 
which  states that the posterior distribution 
$\mathcal{P}({\psi}|\mathcal{D},\mathcal{M})$ of the parameter $\psi$ for the 
model $\mathcal{M}$ with cosmological data set $\mathcal{D}$ can be derived as
  \begin{equation}
  \mathcal{P}({\psi}|\mathcal{D},\mathcal{M}) = \frac{\mathcal{L}(\mathcal{D}|\psi, \mathcal{M}) \pi (\psi|\mathcal{M})}{\mathcal{E}(\mathcal{D}|\mathcal{M})}.
  \end{equation}
Here, $\mathcal{L}(\mathcal{D}|\psi,\mathcal{M})$ is the likelihood of the 
model parameter of $\mathcal{M}$, $\pi (\psi|\mathcal{M})$ is the prior 
probability and $\mathcal{E}(\mathcal{D}|\mathcal{M})$ is the Bayesian 
evidence of the considered cosmological model. The mathematical formulation 
of Bayesian evidence can be written as
  \begin{equation}
\mathcal{E}(\mathcal{D}|\mathcal{M}) = \int_{\mathcal{M}} \mathcal{L}(\mathcal{D}|\psi, \mathcal{M}) \pi (\psi|\mathcal{M}) d\psi,
\end{equation}
The likelihood  $\mathcal{L}(\mathcal{D}|\psi, \mathcal{M})$ has been 
considered as a multivariate Gaussian likelihood function and it has the 
form \cite{akarsu_2019}:
  \begin{equation}
  \mathcal{L}(\mathcal{D}|\psi, \mathcal{M}) \propto \exp\left[\frac{-\,\chi^2(\mathcal{D}|\psi, \mathcal{M})}{2}\right],
  \end{equation}
 in which $\chi^2(\mathcal{D}|\psi, \mathcal{M})$ is the Chi-square function of 
the cosmological data set $\mathcal{D}$. In the case of a uniform prior 
distribution $\pi(\psi|\mathcal{M})$, the posterior distribution can be 
considered as
 \begin{equation}
 \mathcal{P}({\psi}|\mathcal{D},\mathcal{M}) \propto \exp\left[\frac{-\,\chi^2(\mathcal{D}|\psi, \mathcal{M})}{2}\right].
\end{equation}  
We have employed this technique to estimate and constrain various model and 
cosmological parameters with the help of various observational data sets in 
this work.
\subsection{Data and their respective likelihoods}
In this work, we have used observational data of Hubble parameter, BAO, CMB 
and Pantheon supernovae type Ia from various sources and catalogs for 
estimation and constraining of cosmological and model parameters. In the 
following, we have introduced these sets of cosmological data along with 
their respective likelihoods.
\subsubsection{\textbf{Hubble parameter $H(z)$ data}}
For our work, we have collected  57 observational $H(z)$ data from different 
literatures and compiled them in Table \ref{table1}. The chi-square value 
$\chi^2_H$ for the mentioned Hubble data set can be obtained as  
\begin{equation}
\chi^2_{H} =  \sum_{n=1}^{57} \frac{\left[H^{obs}(z_n)- H^{th}(z_n)\right]^2}{\sigma^2_{H^{obs}(z_n)}},
\end{equation}
where $H^{obs}(z_n)$ is the observed Hubble data at the redshift $z_n$, 
$ H^{th}(z_n)$ is the corresponding theoretical Hubble parameter value 
obtained from a considered cosmological model and $\sigma_{H^{obs}(z_n)}$ 
denotes the standard deviation of $n$th observational $H(z)$ data as shown
in Table \ref{table1}.
\begin{center}
\begin{table}[!hbt]
\caption{Available observational Hubble parameter ($H^{obs}(z)$) 
data set in the unit of km/s/Mpc from different literature.}
\vspace{2mm}
\begin{tabular}{ccc|ccc}
\hline 
\rule[-1ex]{0pt}{2.5ex} \hspace{0.5cm} $z$ \hspace{0.5cm}  & \hspace{0.5cm} $ H^{obs}(z)$ \hspace{0.5cm} & \hspace{0.5cm} Reference \hspace{0.5cm} &
\hspace{0.5cm} $z$ \hspace{0.5cm}  & \hspace{0.5cm} $ H^{obs}(z)$ \hspace{0.5cm} & \hspace{0.5cm} Reference \hspace{0.5cm}\\ 
\hline
\rule[-1ex]{0pt}{2.5ex} 0.0708 & 69.0 $\pm$ 19.68 & \cite{Zhang_2014} & 
0.48 & 97.0 $\pm$ 62.0 & \cite{Ratsimbazafy_2017}\\
\rule[-1ex]{0pt}{2.5ex} 0.09 & 69.0 $\pm$ 12.0 & \cite{Simon_2005} &
0.51 & 90.8 $\pm$ 1.9 & \cite{Alam_2017}\\
\rule[-1ex]{0pt}{2.5ex} 0.12 & 68.6 $\pm$ 26.2 & \cite{Zhang_2014} &
0.52 & 94.35 $\pm$ 2.64 & \cite{Wang_2017}\\
\rule[-1ex]{0pt}{2.5ex} 0.17 & 83.0 $\pm$ 8.0 & \cite{Simon_2005} &
0.56 & 93.34 $\pm$ 2.3 & \cite{Wang_2017}\\
\rule[-1ex]{0pt}{2.5ex}0.179 & 75.0 $\pm$ 4.0 & \cite{Moresco_2012} &
0.57 & 92.4 $\pm$ 4.5 & \cite{Samushia_2013} \\
\rule[-1ex]{0pt}{2.5ex} 0.199 & 75.0 $\pm$ 5.0 & \cite{Moresco_2012} &
0.57 & 87.6 $\pm$ 7.8 & \cite{Chuang_2013} \\
\rule[-1ex]{0pt}{2.5ex} 0.20 & 72.9 $\pm$ 29.6 & \cite{Zhang_2014} &
0.59 & 98.48 $\pm$ 3.18 & \cite{Wang_2017}\\
\rule[-1ex]{0pt}{2.5ex} 0.24 & 79.69 $\pm$ 2.65 & \cite{Gaztanaga_2009} &
0.593 & 104.0 $\pm$ 13.0 & \cite{Moresco_2012}\\
\rule[-1ex]{0pt}{2.5ex} 0.27 & 77.0 $\pm$ 14.0 & \cite{Simon_2005} &
0.60 & 87.9 $\pm$ 6.1 & \cite{Blake_2012}\\
\rule[-1ex]{0pt}{2.5ex} 0.28 & 88.8 $\pm$ 36.6 & \cite{Zhang_2014} &
0.61 & 97.8 $\pm$ 2.1 & \cite{Alam_2017}\\
\rule[-1ex]{0pt}{2.5ex} 0.30 & 81.7 $\pm$ 6.22  & \cite{Oka_2014} &
0.64 & 98.82 $\pm$ 2.98 & \cite{Wang_2017}\\
\rule[-1ex]{0pt}{2.5ex} 0.31 & 78.18 $\pm$ 4.74  & \cite{Wang_2017} &
0.6797 & 92.0 $\pm$ 8.0 & \cite{Moresco_2012}\\
\rule[-1ex]{0pt}{2.5ex} 0.34 & 83.8 $\pm$ 3.66  & \cite{Gaztanaga_2009} &
0.73 & 97.3 $\pm$ 7.0 & \cite{Ratsimbazafy_2017}\\
\rule[-1ex]{0pt}{2.5ex} 0.35 & 82.7 $\pm$ 9.1  & \cite{Xu_2013} &
0.781 & 105.0 $\pm$ 12.0 & \cite{Moresco_2012}\\
\rule[-1ex]{0pt}{2.5ex} 0.352 & 83.0 $\pm$ 14.0 & \cite{Moresco_2012}&
0.8754 & 125.0 $\pm$ 17.0 & \cite{Moresco_2012}\\
\rule[-1ex]{0pt}{2.5ex} 0.36 & 79.94 $\pm$ 3.38  & \cite{Wang_2017} &
0.88 & 90.0 $\pm$ 40.0 & \cite{Ratsimbazafy_2017}\\
\rule[-1ex]{0pt}{2.5ex} 0.38 & 81.9 $\pm$ 1.9 & \cite{Alam_2017} &
0.90 & 117.0 $\pm$ 23.0 & \cite{Simon_2005}\\
\rule[-1ex]{0pt}{2.5ex} 0.3802 & 83.0 $\pm$ 13.5 & \cite{Moresco_2016} &
1.037 & 154.0 $\pm$ 20.0 & \cite{Moresco_2012}\\
\rule[-1ex]{0pt}{2.5ex} 0.40 & 82.04 $\pm$ 2.03 & \cite{Wang_2017} &
1.30 & 168.0 $\pm$ 17.0 & \cite{Simon_2005}\\
\rule[-1ex]{0pt}{2.5ex} 0.40 & 95.0 $\pm$ 17.0 & \cite{Simon_2005} &
1.363 & 160.0 $\pm$ 33.6 & \cite{Moresco_2015}\\
\rule[-1ex]{0pt}{2.5ex} 0.4004 & 77.0 $\pm$ 10.2 & \cite{Moresco_2016} &
1.43 & 177.0 $\pm$ 18.0 & \cite{Simon_2005}\\
\rule[-1ex]{0pt}{2.5ex} 0.4247 & 87.1 $\pm$ 11.2 & \cite{Moresco_2016} &
1.53 & 140.0 $\pm$ 14.0 & \cite{Simon_2005}\\
\rule[-1ex]{0pt}{2.5ex} 0.43 & 86.45 $\pm$ 3.68 & \cite{Gaztanaga_2009} &
1.75 & 202.0 $\pm$ 40.0 & \cite{Simon_2005}\\
\rule[-1ex]{0pt}{2.5ex} 0.44 & 82.6 $\pm$ 7.8 & \cite{Blake_2012} &
1.965 & 186.5 $\pm$ 50.4 & \cite{Moresco_2015}\\
\rule[-1ex]{0pt}{2.5ex} 0.44 & 84.81 $\pm$ 1.83 & \cite{Wang_2017} &
2.30 & 224 $\pm$ 8.6 & \cite{Busca_2013}\\
\rule[-1ex]{0pt}{2.5ex} 0.4497 &  92.8 $\pm$ 12.9 & \cite{Moresco_2016} & 
2.33 & 224 $\pm$ 8 & \cite{Bautista_2017}\\
\rule[-1ex]{0pt}{2.5ex} 0.47 & 89.0 $\pm$ 50.0 & \cite{Ratsimbazafy_2017} &
2.34 & 223.0 $\pm$ 7.0 & \cite{Delubac_2015}\\
\rule[-1ex]{0pt}{2.5ex} 0.4783 & 80.9 $\pm$ 9.0 & \cite{Moresco_2016} &
2.36 & 227.0 $\pm$ 8.0 & \cite{Ribera_2014}\\
\rule[-1ex]{0pt}{2.5ex} 0.48 & 87.79 $\pm$ 2.03 & \cite{Wang_2017} &&&\\
\hline
\end{tabular}
\label{table1}
\vspace{0.8cm}
\end{table}
\end{center}

\subsubsection{\textbf{BAO data}}
Baryon acoustic oscillation (BAO) data is associated with the angular diameter 
distance in terms of redshift and it is also useful in studying the 
evolution of $H(z)$. In general, the BAO data provide the dimensionless ratio 
`$d$' of the comoving size of the sound horizon $r_s$ at the drag redshift
 $z_d = 1059.6$ \cite{Planck_2015} to  $D_v (z)$ which is the volume-averaged 
distance. Thus,
\begin{equation}
d = \frac{r_s(z_d)}{D_v (z)},
\end{equation}
where $r_s(z_d)$ is expressed as
\begin{equation}\label{r}
r_s(z_d) = \int^{\infty}_{z_d} \frac{c_s dz}{H(z)}
\end{equation}
and
\begin{equation}\label{DV}
D_v (z) = \left[(1+z)^2 D_A (z)^2 \frac{c\,z}{H(z)}\right]^{\frac{1}{3}}\!\!.
\end{equation}
The term $c_s$ appears in equation \eqref{r} is the sound velocity of the 
baryon-photon fluid with the mathematical expression 
$c_s = c/\sqrt{3(1+\mathcal{R})}$. Here, the term 
$\mathcal{R} = 3 \,\Omega_{b0}/ \left(4\,\Omega_{r0}(1+z)\right)$ in which 
$\Omega_{b0} = 0.022h^{-2}$\cite{Cooke_2016}, $\Omega_{r0} = \Omega_{\gamma 0}\left(1+ 7/8(4/11)^{\frac{4}{3}}N_{eff}\right)$ and $\Omega_{\gamma 0} = 2.469 \times 10^{-5} h^{-2}$ along with $N_{eff} = 3.046$ \cite{Dodelson_2003, akarsu_2019}. Further, $D_A$ in equation \eqref{DV} is the angular diameter distance 
which can be calculated as
\begin{equation}
D_A = \frac{c}{(1+z)} \int^{z}_{0}\!\!\frac{dz}{H(z)},
\end{equation}
where c is the speed of light. We have used 8 BAO data obtained from 
various literatures, which are tabulated in Table \ref{table2} with the 
calculated total standard deviation ($\sigma_{d}^{obs(z_i)}$) for each of them.
\begin{center}
\begin{table}[!hbt]
\caption{Available observational BAO data.}
\vspace{2mm}
\begin{tabular}{ccccc}
\hline 
\rule[-1ex]{0pt}{2.5ex} \hspace{0.5cm} Survey \hspace{0.5cm}&\hspace{0.5cm} $z_i$ \hspace{0.5cm}  & \hspace{0.5cm} $ d^{obs}(z_i)$ \hspace{0.5cm}&\hspace{0.5cm}$\sigma_{d^{obs}(z_i)}$\hspace{0.5cm} & \hspace{0.5cm} Reference \hspace{0.5cm} \\ 
\hline
\rule[-1ex]{0pt}{2.5ex} 6dFGS & 0.106 & 0.3360 & 0.0150 & \cite{Beutler_2011} \\
\rule[-1ex]{0pt}{2.5ex} MGS & 0.15 & 0.2239 & 0.0084 & \cite{Ross_2015} \\
\rule[-1ex]{0pt}{2.5ex} BOSS LOWZ & 0.32 & 0.1181 & 0.0024 & \cite{Padmanabhan_2012} \\
\rule[-1ex]{0pt}{2.5ex} SDSS(R) & 0.35 & 0.1126 & 0.0022 & \cite{Anderson_2014} \\
\rule[-1ex]{0pt}{2.5ex} BOSS CMASS & 0.57 & 0.0726 & 0.0007 & \cite{Padmanabhan_2012} \\
\rule[-1ex]{0pt}{2.5ex} WiggleZ & 0.44 & 0.073 & 0.0012 & \cite{Blake_2012} \\
\rule[-1ex]{0pt}{2.5ex} WiggleZ & 0.6 & 0.0726 & 0.0004 & \cite{Blake_2012} \\
\rule[-1ex]{0pt}{2.5ex} WiggleZ & 0.73 & 0.0592 & 0.0004 & \cite{Blake_2012} \\
\hline
\end{tabular}
\label{table2}
\end{table}
\end{center}

The chi-square value denoted by $\chi^2_{d}$ for the first five data of 
Table \ref{table2} can be computed by using the mathematical expression,
\begin{equation}
\chi^2_{d} = \sum_{i=1}^{5} \frac{\left[d^{obs}(z_i)- d^{th}(z_i)\right]^2}{\sigma^2_{d^{obs}(z_i)}},
\end{equation}
in which $d^{obs}(z_i)$ is the observed value of the dimensionless parameter 
`$d$' at the redshift $z_i$ and $d^{th}(z_i)$ is the corresponding theoretical 
value of `$d$' for a considered cosmological model. For the remaining three 
data of Table \ref{table2} which are taken from WiggleZ survey, the
chi-square value denoted by $\chi^2_w$ can be obtained by using the method 
of covariant matrix. The required inverse of the covariant matrix for the 
considered data set can be obtained from Ref.~\cite{akarsu_2019} as 
given by
 \begin{equation}
C^{-1}_{w}=
\begin{bmatrix}
  1040.3 & -807.5 & 336.8 \\
  -807.5 & 3720.3 & -1551.9\\
  336.8  & -1551.9 & 2914.9
\end{bmatrix}.
\end{equation}
Thus, for the considered three WiggleZ survey data the chi-square value can be 
obtained as
\begin{equation}
\chi^{2}_{w} = D^{T}C^{-1}_wD,
\end{equation}
in which the matrix D has the form:
\begin{equation}
D = 
\begin{bmatrix}
d^{obs}(0.44)- d^{th}(0.44)\\
d^{obs}(0.60)- d^{th}(0.60)\\
d^{obs}(0.73)- d^{th}(0.73)\\
\end{bmatrix}.
\end{equation}
Hence, the total chi-square value for the BAO data set ($\chi^{2}_{BAO}$) of 
Table \ref{table2} can be written as
\begin{equation}
\chi^{2}_{BAO} = \chi^{2}_{d} + \chi^{2}_{w}.
\end{equation}
\subsubsection{\textbf{CMB data}}
The CMB data contain the angular scale of the sound horizon at the last 
scattering surface $l_a$ which is mathematically defined as
\begin{equation}
l_a = \pi\, \frac{r(z^*)}{r_s (z^*)},
\end{equation}
where $r(z_*)$ is the comoving distance to the last scattering surface at
redshift $z^*$ ($= 1089.9$), which can further be defined as
\begin{equation}
r(z_*) = \int^{z^*}_{0}\!\! \frac{c\,dz}{H(z)}.
\end{equation}
Again, the $r_s(z_*)$ is the size of the comoving sound horizon at the 
redshift $z^*$ of the last scattering. The observed value  of 
$l^{obs}_{a} = 301.63 \pm 15$ as per Ref.~\cite{Planck_2015}.

Further, the chi-square value $\chi^2_{CMB}$ for the CMB data can be computed  
as
\begin{equation}
\chi^2_{CMB} = \frac{(l^{obs}_a - l^{th}_a)^2}{\sigma^2_{l_a}},
\end{equation}
where $l^{th}_a$ is the theoretically obtained value for a considered 
model and $\sigma_{l_a}$ is the standard deviation of the observed data 
$l^{obs}_a$.
\subsubsection{\textbf{Pantheon plus supernovae type Ia data}}
\begin{figure}[!h]
\centerline{
  \includegraphics[scale = 0.55]{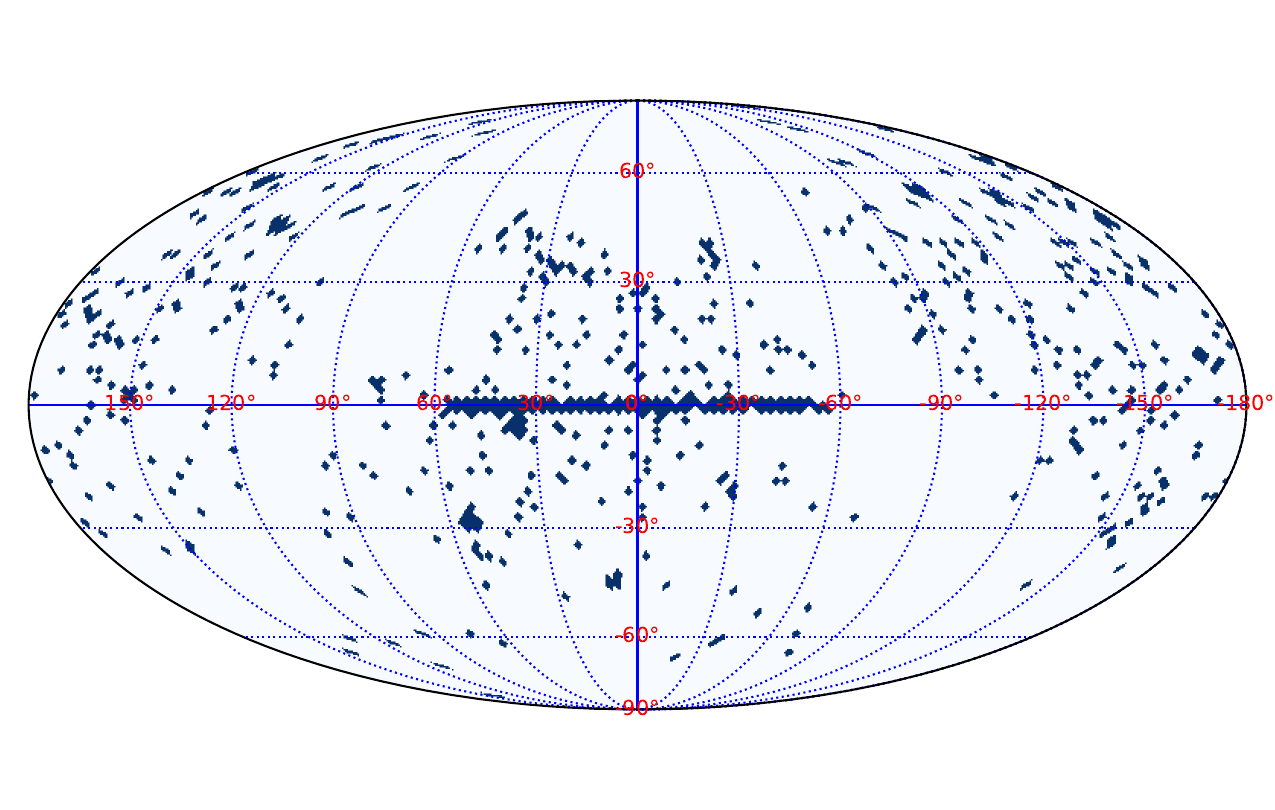}\hspace{0.25cm}
 }
\vspace{-0.2cm}
\caption{Distribution of Type Ia supernovae (SN Ia) in the sky map from 
Pantheon plus data.}
\label{fig1}
\end{figure}

The Pantheon data sample consists of five subsamples PS1, SDSS, SNLS,
low-$z$, and HST \cite{Betoule_2014}. It has the observational data of 1048 
Type Ia supernovae (SN Ia) spanning over the range of $z$ within 
$0.001 < z < 2.3$. The Pantheon plus sample is the updated version of the 
Pantheon sample containing 1701 observational data from 18 different sources 
\cite{Scolnic_2022}. The distribution of these supernovae is shown in 
Fig.~\ref{fig1}. These data compilations contain the information of observed 
peak magnitude $m_B$ and the distance modulus $D_m$ for different SN Ia.

Theoretically, the distance modulus $D_m$ can be calculated by using the 
mathematical expression,
\begin{equation}
D_m = 5 \log_{10} \frac{d_L (z_{hel},z_{cmb})}{10\,\mbox{pc}} = 5 \log_{10} \frac{d_L (z_{hel},z_{cmb})}{1\,\mbox{Mpc}} + 25,
\end{equation}
where the term $z_{hel}$ is the heliocentric redshift, $z_{cmb}$ is the 
redshift of the CMB rest frame and the term $d_L$ is the luminosity distance. 
As discussed in the previous section the theoretical luminosity distance can 
be computed by using the equation \eqref{d_L}. Like in the previously 
mentioned other data set, the chi-square value for the Pantheon plus 
dataset denoted by $\chi^{2}_{Pan+}$ can be obtained by using the covariance 
matrix technique and this can be evaluated as
\begin{equation}
\chi^{2}_{Pan+} = \mathcal{M}^{T}C^{-1}\mathcal{M},
\end{equation}
where $C$ is the total covariance matrix of the observed peak magnitude $m_B$ 
and $\mathcal{M} = m_B - m_B^{th}$ with 
\begin{equation}
m_B^{th} = 5 \log_{10} D_{L} + M.
\end{equation}
Here,
\begin{equation}
D_{L} = (1+z_{hel}) \int^{Z_{cmb}}_{0} \frac{H_0\, dz}{H(z)}.
\end{equation}
and the term $M$ is the nuisance parameter. For the Pantheon data set, the 
value of $M$ is $23.739^{+0.140}_{-0.102}$ \cite{Zhao_2019}. Moreover, the 
total covariance matrix $C$ can be expressed as 
 \begin{equation}
 C = C_{sys} + C_{ds},
 \end{equation}
where $C_{sys}$ consists of a systematic covariance matrix and $C_{ds}$ is the 
diagonal covariance matrix of the statistical uncertainty 
\cite{Scolnic_2018, akarsu_2019}.

\subsection{Constraining of cosmological parameters}
\subsubsection{$f(R,T) = \alpha R + \beta \lambda T$}
\begin{figure}[!h]
\centerline{
  \includegraphics[scale = 0.55]{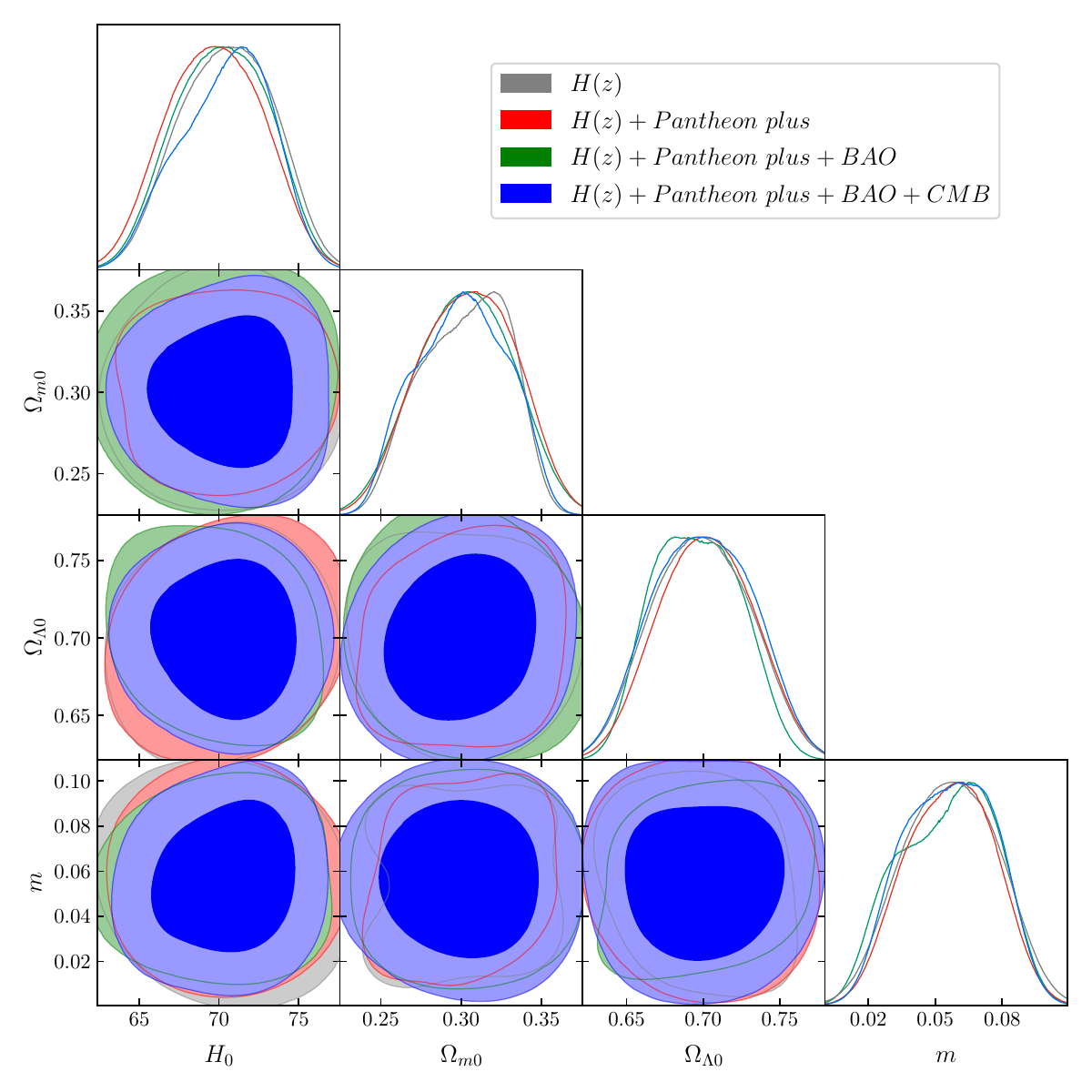}\hspace{0.25cm}
 }
\vspace{-0.2cm}
\caption{One-dimensional and two-dimensional marginalized confidence regions 
($68\%$ and $95\%$ confidence levels) of cosmological and model parameters for 
the anisotropic $f(R,T)$-I BIII model obtained with the help of $H(z)$, 
Pantheon plus, BAO and CMB data.}
\label{fig2}
\end{figure}

For the convenience of the representation we named this model of $f(R,T)$ 
gravity along with the BIII metric as anisotropic $f(R,T)$-I BIII model of 
the Universe. To implement observational constraints on this 
anisotropic $f(R,T)$-I BIII model, we have taken a multivariate 
joint Gaussian likelihood of the form \cite{akarsu_2019}:
\begin{equation}\label{L}
\mathcal{L}_{tot} \propto \exp\left(\frac{-\,\chi^2_{tot}}{2}\right),
\end{equation}
where 
\begin{equation}
\chi^2_{tot} = \chi^2_{H} + \chi^2_{BAO} +\chi^2_{CMB}+\chi^2_{Pan+}
\end{equation}
Here, we have considered uniform prior distributions for all cosmological
parameters as well as for model parameters of the considered anisotropic 
$f(R,T)$-I BIII model. 
The prior ranges of various parameters have been considered as follows: 
$55 < H_0 < 85$, $0.1 < \Omega_{mo} < 0.5$, $0.00001 < \Omega_{ro}  < 0.0001$, 
$0.6 < \Omega_{\Lambda 0} < 1$, $0.001< m <0.01$, $0.95< \alpha <1.05$, 
$1.5< \beta <2.5$, $0.01 < \lambda <0.1$, $0.95 < \gamma < 1.05$. The 
likelihoods are considered within these mentioned ranges such that results 
should be consistent with standard Planck data release 2018 \cite{Planck_2018} along with
the current observational data. With these considerations, we have plotted 
one-dimensional and two-dimensional marginalized confidence regions 
($68\%$ and $95\%$ confidence levels) for the anisotropic $f(R,T)$-I BIII 
model, in which we mainly focused on cosmological parameters like $H_0$, 
$\Omega_{mo}$, $\Omega_{\Lambda0}$ etc.~along with the estimation of the model 
parameters like  $m$ and $\alpha$, $\beta$, $\gamma$ and $\lambda$ for $H(z)$, 
 $H(z)$ + Pantheon plus, $H(z)$ + Pantheon plus + BAO and 
$H(z)$ + Pantheon plus + BAO + CMB data sets as shown in Fig.~\ref{fig2}.

\begin{figure}[!h]
\centerline{
  \includegraphics[scale = 0.5]{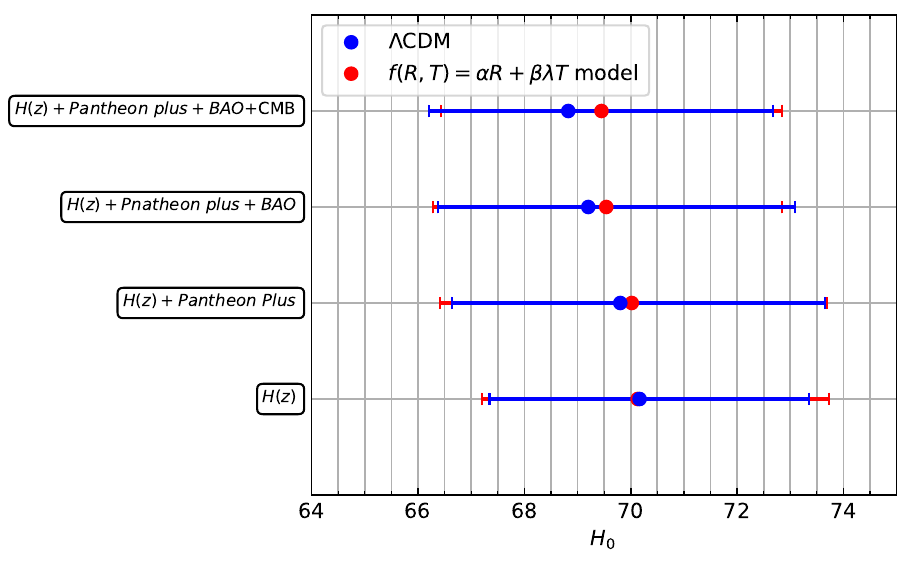}\hspace{0.05cm}
  \includegraphics[scale = 0.5]{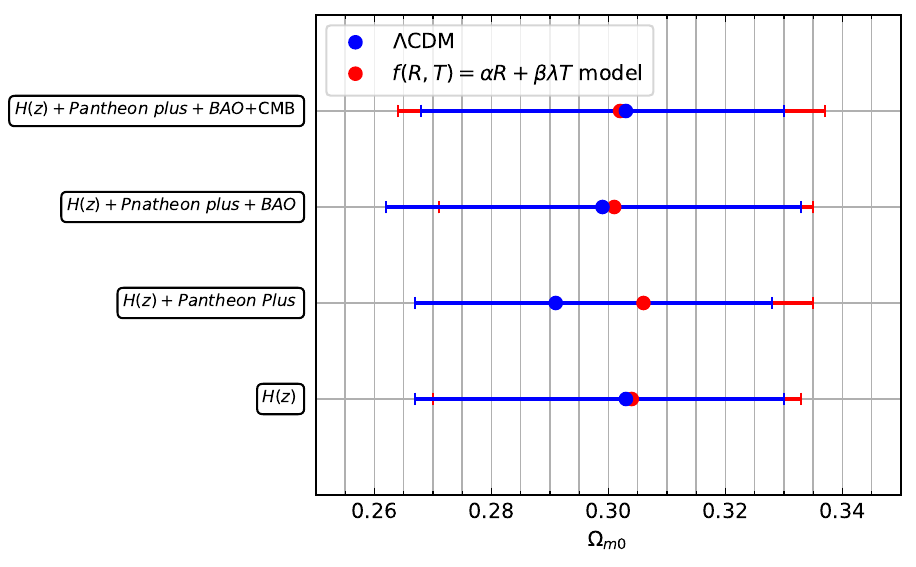}
 }
 \vspace{5pt}
 \centerline{
  \includegraphics[scale = 0.5]{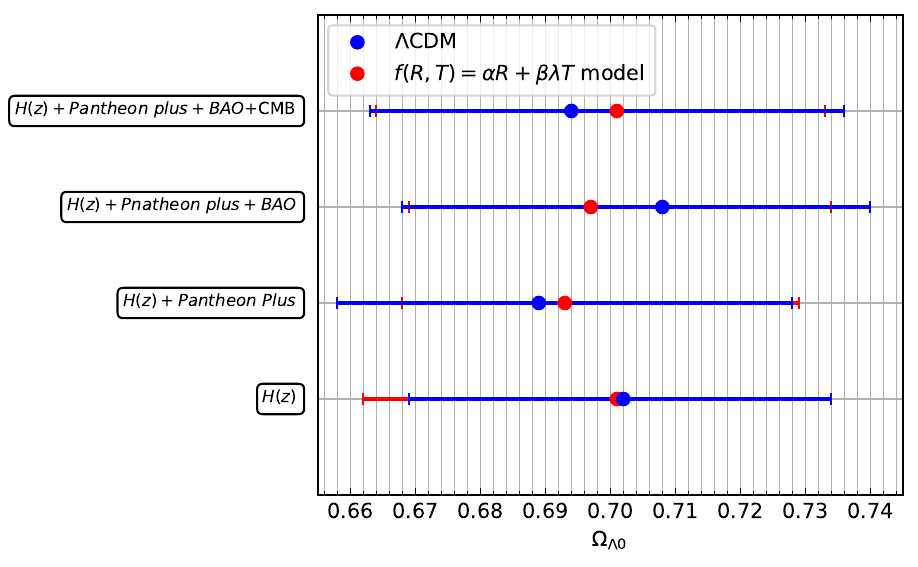}\hspace{0.05cm}
  \includegraphics[scale = 0.5]{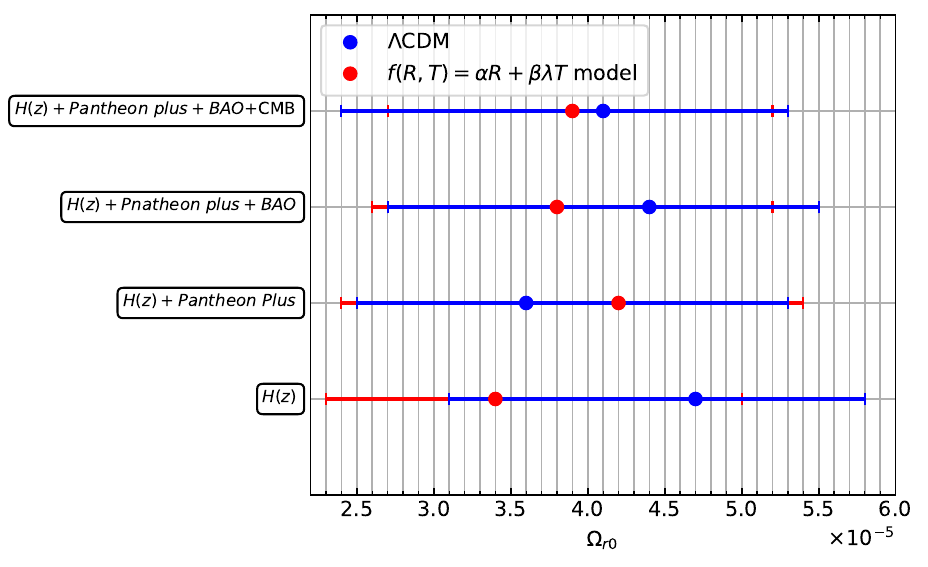}
 }
\vspace{-0.2cm}
\caption{$68\%$ confidence level intervals of $H_0$, $\Omega_{m0}$, 
$\Omega_{\Lambda 0}$ and $\Omega_{r0}$ for the anisotropic $f(R,T)$-I BIII 
model in comparison with that of the $\Lambda$CDM model.}
\label{fig3}
\end{figure}

Table \ref{table3} shows the constraints ($68\%$ confidence level) 
on the model parameters and cosmological parameters for the anisotropic
$f(R,T)$-I BIII model and $\Lambda$CDM model obtained from different 
available 
data sets as mentioned above. From Table \ref{table3} and Fig.~\ref{fig2}, we 
found that the tightest constraint can be obtained from the joint data set of 
$H(z)$ + Pantheon plus + BAO + CMB on a maximum number of the parameters for 
both the anisotropic $f(R,T)$-I BIII model and $\Lambda$CDM model.  
\begin{center}
\begin{table}[!h]
\caption{Constrained values of cosmological parameters including model-specific
parameters for both anisotropic $f(R,T)$-I BIII model and $\Lambda$CDM model 
obtained through $68\%$ confidence level corner plots using different 
cosmological data sources.}
\vspace{8pt}
\scalebox{0.88}{
\begin{tabular}{cccccc}
\hline 
\rule[0.1ex]{0pt}{2.5ex} Model \hspace{2pt} & \hspace{2pt} Parameters \hspace{2pt} & \hspace{2pt} $H(z)$ \hspace{2pt} & \hspace{2pt} $H(z) + \text{Pantheon plus}$ \hspace{2pt} & \hspace{2pt} $H(z) + \text{Pantheon plus} + \text{BAO}$ \hspace{2pt} &\hspace{2pt} $H(z) + \text{Pantheon plus} + \text{BAO} + \text{CMB}$ \hspace{0.25cm}\\
\vspace{-9.5pt}\\
\hline\\[-8pt]
\rule[-2ex]{0pt}{2.5ex}&$H_0$& ${ 70.133^{+3.600}_{-2.928}}$ & ${ 70.020^{+3.667}_{-3.601}}$& ${ 69.535^{+3.297}_{-3.262}}$&${ 69.450^{+3.400}_{-3.015}}$\\
\rule[-2ex]{0pt}{2.5ex}&$\Omega_{m0}$ &${ 0.304^{+0.029}_{-0.034}}$ & ${ 0.306^{+0.029}_{-0.039}}$ &${ 0.301^{+0.034}_{-0.030}}$&${ 0.302^{+0.035}_{-0.038}}$\\
\rule[-2ex]{0pt}{2.5ex}&$\Omega_{r0}$ &${ 0.000034^{+0.000016}_{-0.000011}}$ & ${ 0.000042^{+0.000012}_{-0.000018}}$ &${ 0.000038^{+0.000014}_{-0.000012}}$& ${ 0.000039^{+0.000013}_{-0.000012}}$\\
\rule[-2ex]{0pt}{2.5ex}&$\Omega_{\Lambda0}$&${ 0.701^{+0.033}_{-0.039}}$&$ { 0.693^{+0.036}_{-0.025}}$&${ 0.697^{+0.037}_{-0.028}}$&${ 0.701^{+0.032}_{-0.037}}$\\
\rule[-2ex]{0pt}{2.5ex} $f(R,T)$-I BIII & $m$&${ 0.054^{+0.026}_{-0.020}}$&${ 0.059^{+0.029}_{-0.030}}$ & ${ 0.060^{+0.020}_{-0.023}}$ &${ 0.055^{+0.023}_{-0.025}}$\\
\rule[-2ex]{0pt}{2.5ex}&$\alpha$ &${ 0.998^{+0.018}_{-0.012}}$ &${ 1.001^{+0.014}_{-0.013}}$ &${ 1.002^{+0.012}_{-0.013}}$&${ 1.001^{+0.011}_{-0.015}}$ \\
\rule[-2ex]{0pt}{2.5ex}&$\beta$ &${ 1.982^{+0.197}_{-0.170}}$ & ${ 2.017^{+0.154}_{-0.204}} $ &${ 2.000^{+0.203}_{-0.197}}$&${ 1.998^{+0.215}_{-0.151}}$ \\
\rule[-2ex]{0pt}{2.5ex}&$\gamma$ &${ 0.992^{+0.016}_{-0.013} }$&${ 0.995^{+0.017}_{-0.019}} $ &${ 0.994^{+0.018}_{0.018}}$&${ 0.996^{+0.021}_{-0.018}}$\\
\rule[-2ex]{0pt}{2.5ex}&$\lambda$ &${ 0.044^{+0.012}_{-0.010}}$ &${ 0.044^{+0.011}_{-0.011}}$&${ 0.044^{+0.011}_{-0.009}}$&${ 0.045^{+0.011}_{-0.013}}$\\
\hline\\[-8pt]
\rule[-2ex]{0pt}{2.5ex}&$H_0$&$70.167^{+3.192}_{-2.823}$ & $69.804^{+3.841}_{-3.169}$&$69.202^{+3.893}_{-2.833}$&$68.826^{+3.857}_{-2.620}$\\
\rule[-2ex]{0pt}{2.5ex}$\Lambda$CDM & $\Omega_{m0}$ &$0.303^{+0.027}_{-0.036}$&$0.291^{+0.037}_{-0.024}$ &$0.299^{+0.034}_{-0.037}$&$0.303^{+0.027}_{-0.035}$\\
\rule[-2ex]{0pt}{2.5ex}&$\Omega_{r0}$ &$0.000047^{+0.000011}_{-0.000016}$ & $0.000036^{+0.000017}_{-0.000011}$ &$0.000044^{+0.000011}_{-0.000017}$&$0.000041^{+0.000012}_{-0.000017}$\\
\rule[-2ex]{0pt}{2.5ex}& $\Omega_{\Lambda0}$&$0.702^{+0.032}_{-0.033}$& $0.689^{+0.039}_{-0.031}$ &$0.708^{+0.032}_{-0.040}$&$0.694^{+0.042}_{-0.031}$\\
\hline
\end{tabular}}
\label{table3}
\end{table}
\end{center}
With the use of Table \ref{table3}, we have tried to compare the values of the 
$H_0$, $\Omega_{m0}$, $\Omega_{\Lambda0}$ and $\Omega_{r0}$  parameters for 
both the models for different data sets' combinations within $68\%$ confidence 
intervals as shown in Fig.~\ref{fig3}. The shift of the parameter values from 
the standard $\Lambda$CDM due to an anisotropic background is clearly observed 
in these plots. The largest deviations of the cosmological parameters as seen
from these plots are compiled in Table \ref{table4} for both the standard 
$\Lambda$CDM model and the $f(R,T)$-I BIII anisotropic cosmological model. From 
this table, we can conclude that the deviations are higher in the $\Lambda$CDM 
model in comparison to that of the anisotropic $f(R,T)$-I BIII model.

\begin{center}
\begin{table}[!h]
\caption{Deviations of values of cosmological parameters for different 
combinations of data sets for the $\Lambda$CDM model and the anisotropic 
$f(R,T)$-I BIII model.}
\vspace{8pt}
\scalebox{1}{
\begin{tabular}{ccccc}
\hline 
\rule[0.1ex]{0pt}{2.5ex}Model \hspace{0.25cm} &\hspace{0.25cm} $\Delta H_0$ \hspace{0.25cm}  & \hspace{0.25cm} $\Delta \Omega_{m0} $ \hspace{0.25cm} & \hspace{0.25cm} $\Delta \Omega_{\Lambda0}$ \hspace{0.25cm} &\hspace{0.25cm} $\Delta \Omega_{r0}$ \hspace{0.25cm}\\ 
\hline
\rule[0.1ex]{0pt}{2.5ex}$f(R,T)$-I BIII &{ 0.683} &{0.005} &{0.008} &{0.000005}\\
\rule[0.1ex]{0pt}{2.5ex}$\Lambda$CDM& 1.341  &0.012&0.008  &0.000011\\
\hline
\end{tabular}}
\label{table4}
\end{table}
\end{center}

\begin{figure}[!h]
\centerline{
  \includegraphics[scale = 0.65]{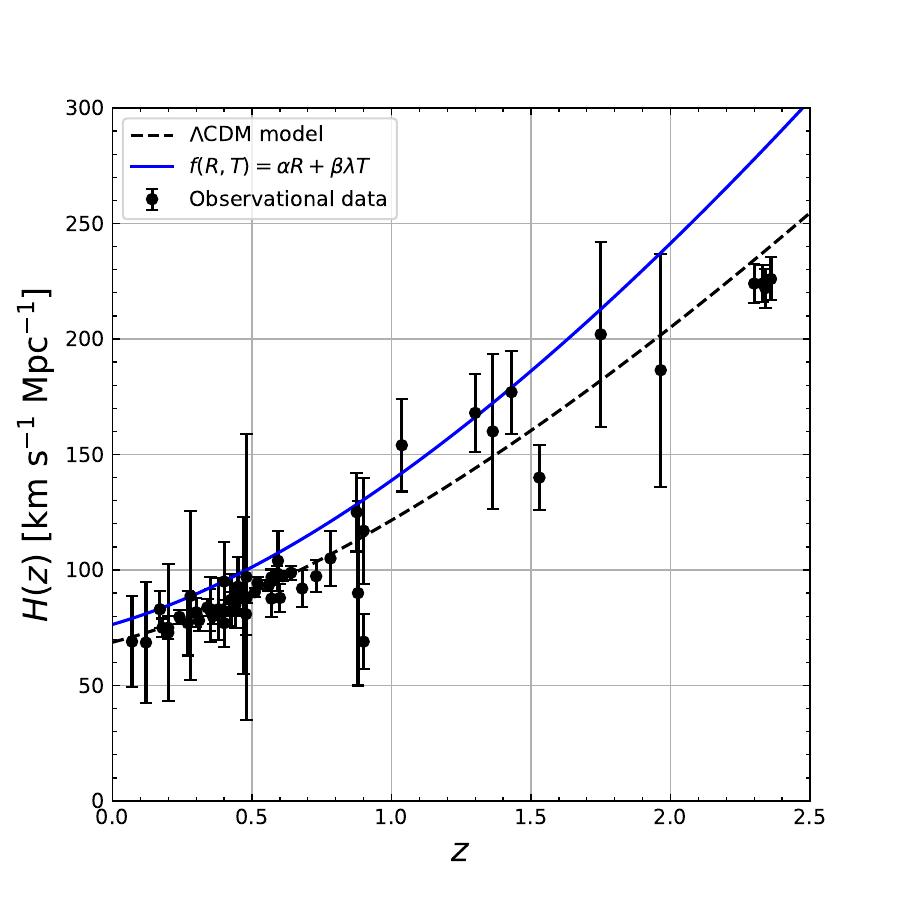}\hspace{0.25cm}
 }
\vspace{-0.3cm}
\caption{Variation of Hubble parameter $H(z)$ against cosmological redshift
$z$ for the constrained set of model parameters of both $\Lambda$CDM and 
anisotropic $f(R,T)$-I BIII models in comparison with the observational data.}
\label{fig4}
\end{figure}

\begin{figure}[!h]
\centerline{
  \includegraphics[scale = 0.6]{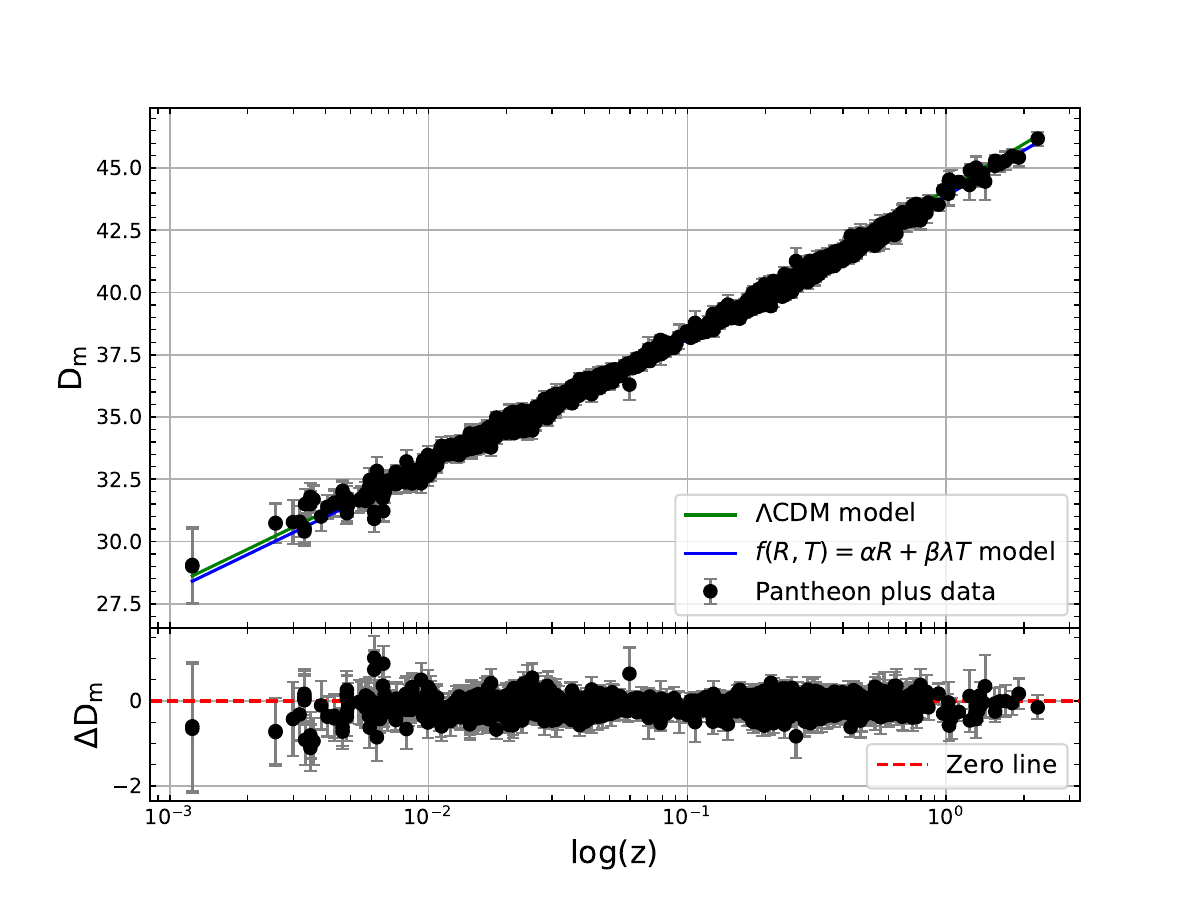}\hspace{0.25cm}
 }
\vspace{-0.2cm}
\caption{Top panel: The Pantheon plus ``Hubble diagram" showing the distance 
modulus $D_m$ versus log of cosmological redshift $z$ for the anisotropic 
$f(R,T)$-I BIII model in comparison with 
the $\Lambda$CDM results. Bottom panel: Distance modulus residues against 
cosmological redshift for Pantheon plus data relative to anisotropic $f(R,T)$-I
BIII model of the Universe.}
\label{fig5}
\end{figure}

Moreover, we have tried to compare the Hubble parameter versus cosmological 
redshift variations for both the models taking the parameters constrained using
the combination of $H(z)$ + Pantheon plus + BAO + CMB data from Table 
\ref{table3} as shown in Fig.~\ref{fig4}. The plot shows that for the 
estimated values of cosmological parameters, the Hubble parameter is 
consistent with the observational data. However, the anisotropic $f(R,T)$-I 
BIII model shows deviations from the
standard $\Lambda$CDM plots with the increase of cosmological redshift $z$. 
From Fig.~\ref{fig4}, we have found that the expansion rate of the 
anisotropic $f(R,T)$-I BIII model is higher in comparison to the 
standard $\Lambda$CDM model as the redshift value $z$ increases. Similarly, we 
have plotted the distance modulus $D_m$ against cosmological redshift
$z$ in Fig.~\ref{fig5} for both $\Lambda$CDM model and anisotropic 
$f(R,T)$-I BIII model along with distance modulus residues 
relative to BIII Universe in the logarithmic $z$ scale for the constrained set 
of model parameters as mentioned above for the $H(z)$ versus $z$ plot. The 
plot shows that like the $\Lambda$CDM model, the distance modulus for the 
anisotropic $f(R,T)$-I BIII model is consistent with the 
observational Pantheon plus data obtained from different SN Ia for the 
constrained set of model parameters of Table \ref{table3}. Further, the plot 
of the distance modulus residues also shows that the model is consistent with 
observational data.
\begin{figure}[!h]
\centerline{
  \includegraphics[scale = 0.45]{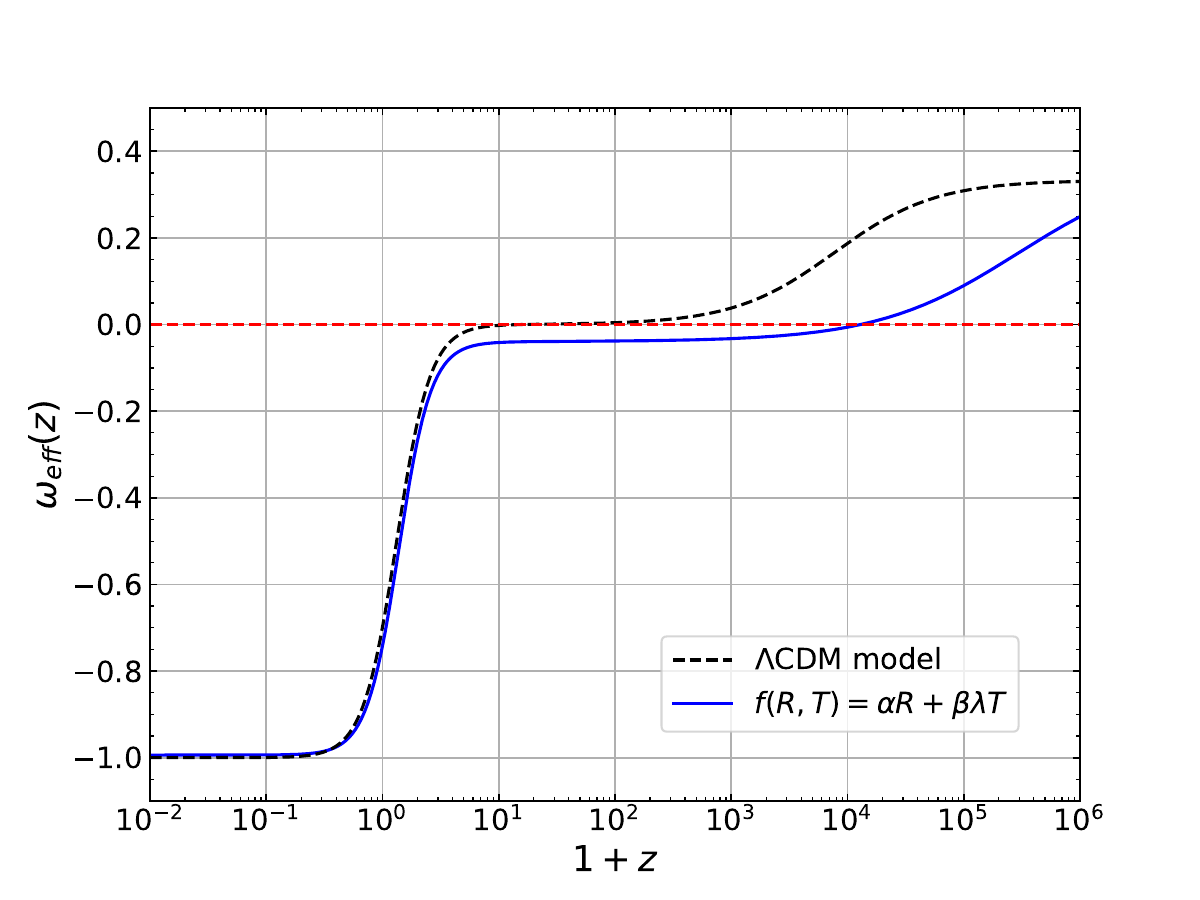}\hspace{0.0cm}
  \includegraphics[scale = 0.45]{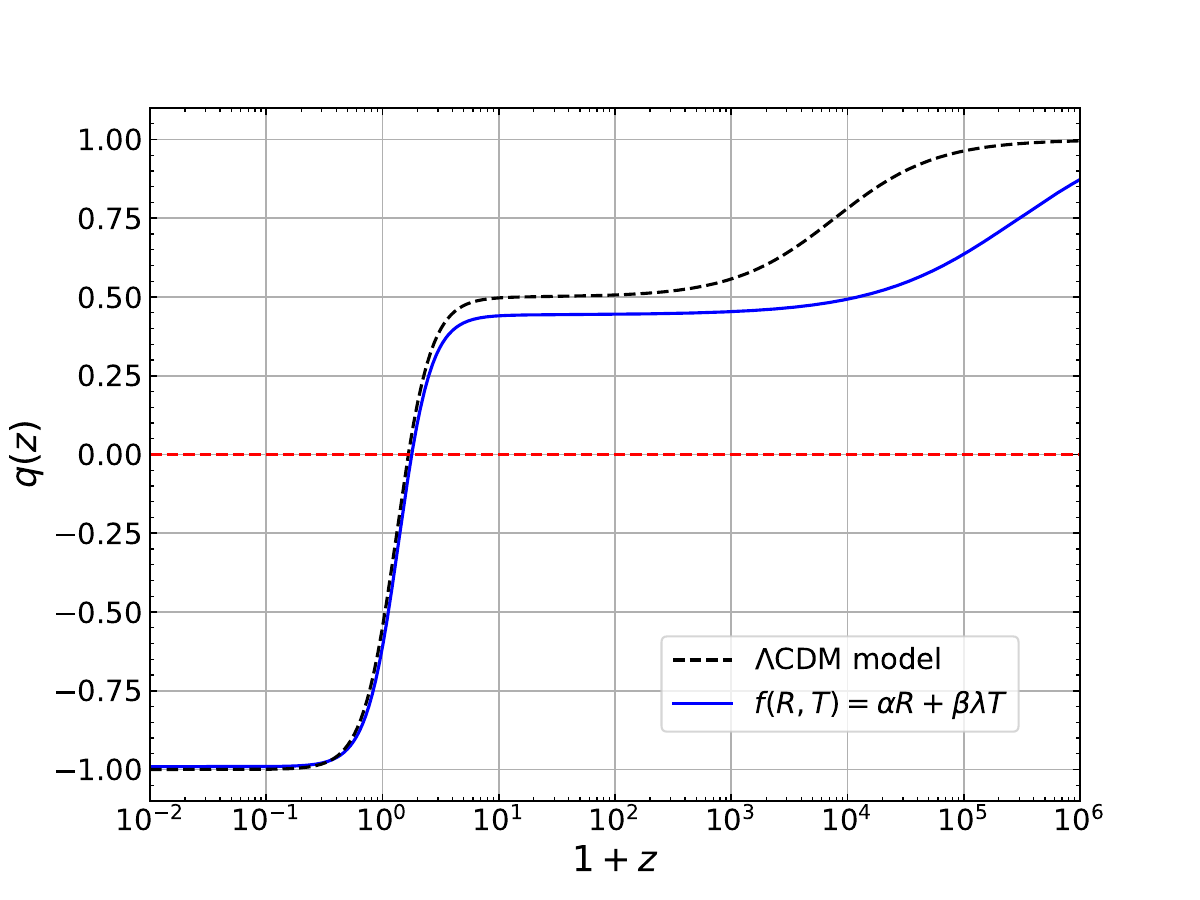}
 }
\vspace{-0.2cm}
\caption{Variation of the effective equation of state $\omega_{eff}$ (left) 
and deceleration parameter $q$ (right) against cosmological redshift $z$ for 
a constrained set of parameters for both the anisotropic $f(R,T)$-I BIII model 
and the $\Lambda$CDM model.}
\label{fig6}
\end{figure}

Furthermore, we have plotted the effective equation of state $\omega_{eff}$ 
using equation \eqref{omf_new1} against cosmological redshift $z$ for the 
constrained set of parameters mentioned in Table \ref{table3} for both 
$\Lambda$CDM model and the considered anisotropic $f(R,T)$-I BIII model in 
Fig.~\ref{fig6} (left). The plot shows the deviation of the anisotropic 
$f(R,T)$-I BIII model results from the standard 
$\Lambda$CDM results for values of $z>0$, indicating the role of anisotropy 
in the matter-dominated and radiation-dominated phases of the Universe. 
Moreover, the deviation between these two models is also seen for values of 
$z<0$, implying that anisotropy will play a role in the future of the Universe. 
Apart from the $\omega_{eff}$, we have also plotted the deceleration parameter 
$q$ from equation \eqref{dec} against cosmological redshift $z$ in 
Fig.~\ref{fig6} (right) for both anisotropic BIII Universe for the considered 
$f(R,T)$ model and standard $\Lambda$CDM model. The anisotropic BIII Universe 
in the considered $f(R,T)$ model $q$ plot also shows deviations from the 
standard $\Lambda$CDM results for the values of $z>0$ and $z<0$ as shown in 
the case of $\omega_{eff}$ plot, again indicating the role of anisotropy in the 
evolution of the Universe. However, both plots show that they are consistent 
with the standard cosmology for $z = 0$, i.e.~in the present time. Thus we can 
comment here that the $f(R,T) = \alpha R + \beta f(T)$ with the considered 
$f(T) = \lambda T$ is a physically viable model to explain the anisotropic 
Universe.

\subsubsection{$f(R,T) = R + 2 \lambda T$}
\begin{figure}[!h]
\centerline{
  \includegraphics[scale = 0.55]{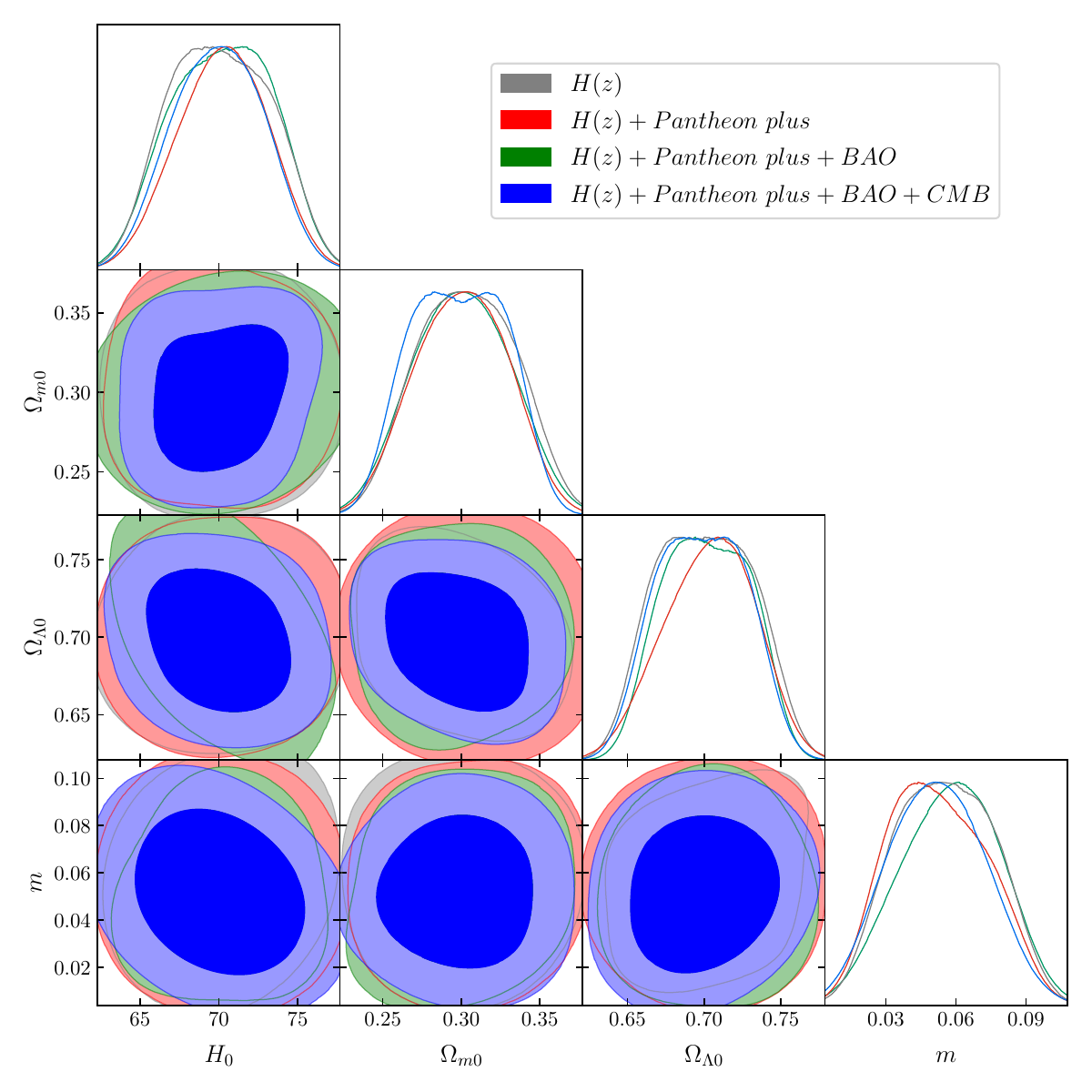}
 }
\vspace{-0.2cm}
\caption{One-dimensional and two-dimensional marginalized confidence regions 
($68\%$ and $95\%$ confidence levels) of cosmological and model parameters for 
the $f(R,T)$-II BIII model obtained with the help of $H(z)$, Pantheon plus, 
BAO and CMB data.}
\label{fig7}
\end{figure}

As in the previous case, for the ease of naming this $f(R,T)$ gravity model
along with the anisotropic BIII metric, we call it the anisotropic 
$f(R,T)$-II BIII model of the Universe. Also like in the previous case, we 
have considered the Gaussian likelihood as equation \eqref{L}. The prior 
ranges of the cosmological parameters and model parameters for this considered 
model are taken as $55 < H_0 < 85$, 
$0.1 <\Omega_{m0} < 0.5$, $0.00001 < \Omega_{r0}  < 0.0001$, 
$0.6 < \Omega_{\Lambda 0}< 1$, $0.01< m <0.1$, $0.01 < \lambda <0.1$,
$0.95 < \gamma < 1.05 $. With these prior ranges, we have plotted 
one-dimensional and two-dimensional marginalized confidence regions 
($68\%$ and $95\%$ confidence levels) for the anisotropic $f(R,T)$-II BIII 
model, in which we mainly focused on 
cosmological parameters like $H_0$, $\Omega_{m0}$, $\Omega_{\Lambda0}$ 
etc.~along with the model parameters like $m$, $\alpha$, $\beta$, $\gamma$ and 
$\lambda$, within the range of $H(z)$, $H(z)$ + Pantheon~plus, $H(z)$ + 
Pantheon plus + BAO and $H(z)$ + Pantheon plus + BAO + CMB data as shown in 
Fig.~\ref{fig7}. In a similar line, we have compiled Table \ref{table5} which 
shows the constraints ($ 68\%$ confidence level) on the considered $f(R,T)$-II 
BIII model and the $\Lambda$CDM model parameters from the 
different available data sets by using the Bayesian inference technique. From 
Table \ref{table5} and Fig.~\ref{fig7}, we have found that like in the 
previous case, the tightest constraint can be achieved from the joint dataset 
of $H(z)$ + Pantheon plus + BAO + CMB on all the cosmological parameters for 
both the anisotropic $f(R,T)$-II BIII model and the $\Lambda$CDM model. 

\begin{figure}[!h]
\centerline{
  \includegraphics[scale = 0.5]{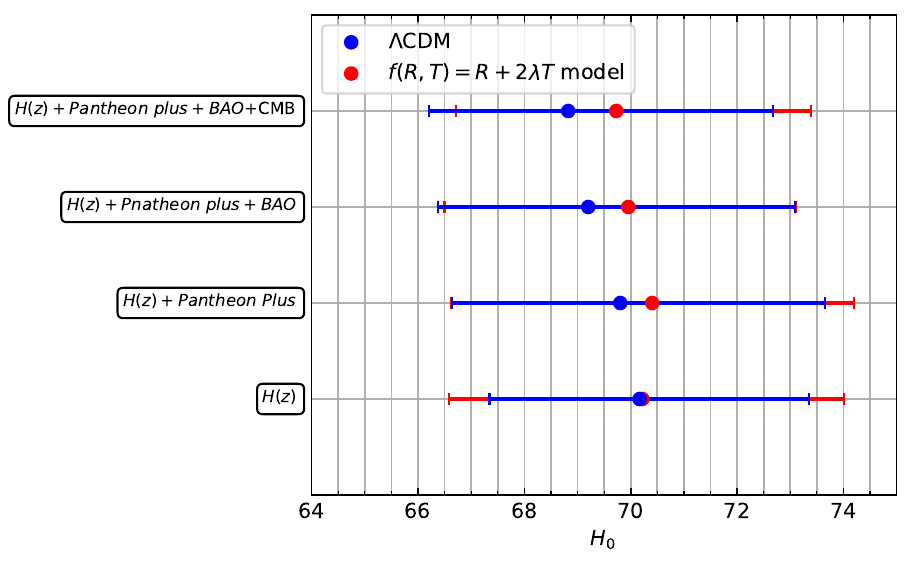}\vspace{0.05cm}
  \includegraphics[scale = 0.5]{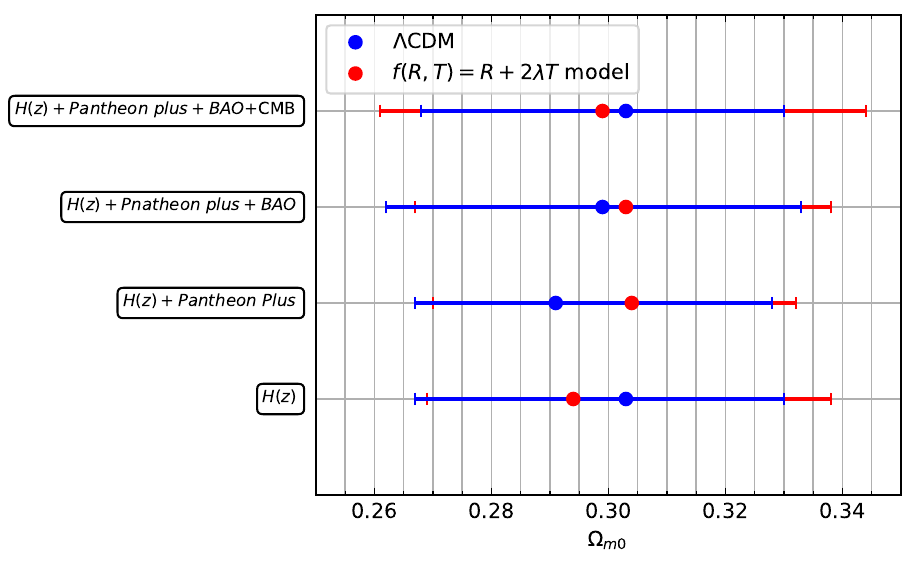}
 }
 \centerline{
  \includegraphics[scale = 0.5]{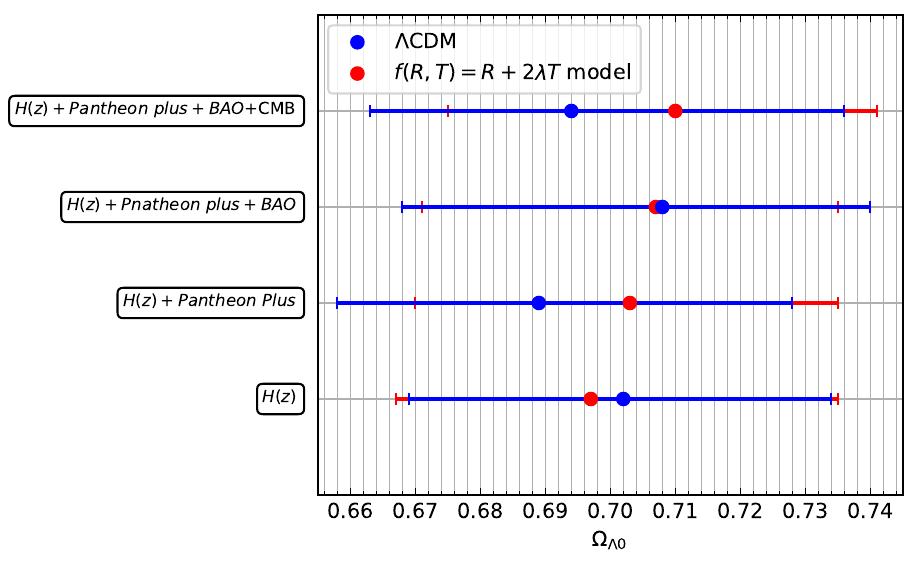}\vspace{0.05cm}
  \includegraphics[scale = 0.5]{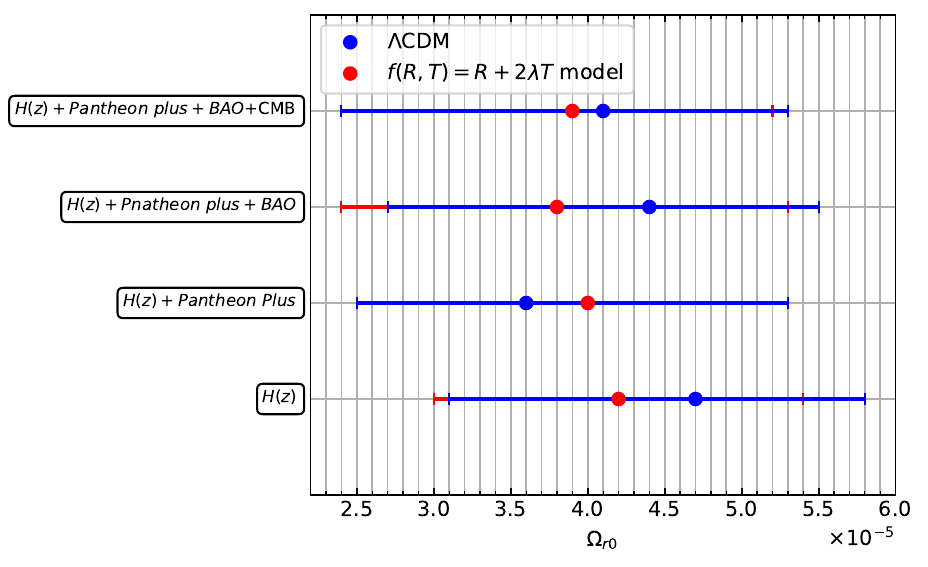}
 }
\vspace{-0.2cm}
\caption{$68\%$ confidence intervals of $H_0$, $\Omega_{m0}$, 
$\Omega_{\Lambda 0}$ and $\Omega_{r0}$ for the anisotropic $f(R,T)$-II BIII 
model in comparison with that of the $\Lambda$CDM model.}
\label{fig8}
\end{figure}

With the use of Table \ref{table5}, we have tried to compare the $H_0$,
$\Omega_{m0}$, $\Omega_{\Lambda0}$ and $\Omega_{r0}$ parameters for both the
models for different data set combinations within the $68\%$ confidence
interval as shown in Fig.~\ref{fig8}. Like in the previous case, the shift
of the parameter values from the standard $\Lambda$CDM model due to the
anisotropic background is clearly observed from the plots of this figure. The
largest deviations of the cosmological parameters from the above plots are
compiled in Table \ref{table6} for both the standard $\Lambda$CDM model and the
$f(R,T)$-II BIII model. From this Table, we can conclude that
the deviations are higher in the $\Lambda$CDM model in comparison to the
anisotropic $f(R,T)$-II BIII model.
\begin{center}
\begin{table}[!h]
\caption{Constrained values of cosmological parameters including model-specific parameters for both the anisotropic $f(R,T)$-II BIII model and the 
$\Lambda$CDM model obtained through confidence level corner plots using 
different cosmological data sources.}
\vspace{5pt}
\scalebox{0.88}{
\begin{tabular}{cccccc}
\hline 
\rule[1ex]{0pt}{2.5ex} Model \hspace{2pt} &\hspace{2pt} Parameters \hspace{2pt} & \hspace{2pt} $H(z)$ \hspace{2pt}  & \hspace{2pt} $H(z)$ + Pantheon plus \hspace{2pt} & \hspace{2pt} $H(z)$ + Pantheon plus + BAO \hspace{2pt} & \hspace{2pt} $H(z)$ + Pantheon~ plus + BAO + CMB \hspace{0.25cm}\\ 
\vspace{-9.5pt}\\
\hline\\[-8pt]
\rule[-2ex]{0pt}{2.5ex}&$H_0$& ${ 70.219^{+3.795}_{-3.632}}$ & ${ 70.404^{+3.800}_{-3.776}}$& ${ 69.955^{+3.150}_{-3.457}}$&${ 69.728^{+3.664}_{-3.019}}$\\
\rule[-2ex]{0pt}{2.5ex}&$\Omega_{m0}$ &${ 0.294^{+0.044}_{-0.025}}$ & ${ 0.304^{+0.028}_{-0.034}}$ &${ 0.303^{+0.035}_{-0.036}}$&${ 0.299^{+0.045}_{-0.038}}$\\
\rule[-2ex]{0pt}{2.5ex}&$\Omega_{r0}$ &${ 0.000042^{+0.000012}_{-0.000012}}$ & ${ 0.000040^{+0.000013}_{-0.000015}}$ &${ 0.000038^{+0.000015}_{-0.000014}}$& ${ 0.000039^{+0.000013}_{-0.000015}}$\\
\rule[-2ex]{0pt}{2.5ex}&$\Omega_{\Lambda0}$&${ 0.697^{+0.038}_{-0.030}}$&$ { 0.703^{+0.033}_{-0.032}}$&${ 0.707^{+0.028}_{-0.036}}$&${ 0.710^{+0.031}_{-0.035}}$\\
\rule[-2ex]{0pt}{2.5ex} $f(R,T)$-II BIII&$m$&${ 0.052^{+0.028}_{-0.024}}$&${  0.055^{+0.025}_{-0.023}}$ & ${ 0.057^{+0.022}_{-0.025}}$ &${ 0.055^{+0.023}_{-0.022}}$\\
\rule[-2ex]{0pt}{2.5ex}&$\gamma$ &${ 0.997^{+0.016}_{-0.019} }$&${ 0.998^{+0.014}_{-0.017} }$ &${ 0.996^{+0.014}_{-0.017}}$&${ 0.997^{+0.018}_{-0.016}}$\\
\rule[-2ex]{0pt}{2.5ex}&$\lambda$ &${ 0.047^{+0.006}_{-0.013}}$ &${ 0.043^{+0.010}_{-0.010}}$&${ 0.044^{+0.015}_{-0.011}}$&${ 0.046^{+0.009}_{-0.012}}$\\
\hline\\[-8pt]
\rule[-2ex]{0pt}{2.5ex}&$H_0$&$70.167^{+3.192}_{-2.823}$ & $69.804^{+3.841}_{-3.169}$&$69.202^{+3.893}_{-2.833}$&$68.826^{+3.857}_{-2.620}$\\
\rule[-2ex]{0pt}{2.5ex}$\Lambda$CDM&$\Omega_{m0}$ &$0.303^{+0.027}_{-0.036}$&$0.291^{+0.037}_{-0.024}$ &$0.299^{+0.034}_{-0.037}$&$0.303^{+0.027}_{-0.035}$\\
\rule[-2ex]{0pt}{2.5ex}&$\Omega_{r0}$ &$0.000047^{+0.000011}_{-0.000016}$ & $0.000036^{+0.000017}_{-0.000011}$ &$0.000044^{+0.000011}_{-0.000017}$&$0.000041^{+0.000012}_{-0.000017}$\\
\rule[-2ex]{0pt}{2.5ex}&$\Omega_{\Lambda0}$&$0.702^{+0.032}_{-0.033}$& $0.689^{+0.039}_{-0.031}$ &$0.708^{+0.032}_{-0.040}$&$0.694^{+0.042}_{-0.031}$\\
\hline
\end{tabular}}
\label{table5}
\end{table}
\end{center}

\begin{center}
\begin{table}[!h]
\caption{Deviations of cosmological parameters for different combinations of 
data sets for the $\Lambda$CDM model and the anisotropic $f(R,T)$-II BIII 
model of the Universe.}
\vspace{8pt}
\scalebox{1}{
\begin{tabular}{ccccc}
\hline 
\rule[0.1ex]{0pt}{2.5ex}Model \hspace{0.25cm} &\hspace{0.25cm} $\Delta H_0$ \hspace{0.25cm}  & \hspace{0.25cm} $\Delta \Omega_{m0} $ \hspace{0.25cm} & \hspace{0.25cm} $\Delta \Omega_{\Lambda0}$ \hspace{0.25cm} &\hspace{0.25cm} $\Delta \Omega_{r0}$ \hspace{0.25cm}\\ 
\hline
\rule[0.1ex]{0pt}{2.5ex}$f(R,T)$-II BIII & {0.676} &{0.010} &{0.013} &{0.000004}\\
\rule[0.1ex]{0pt}{2.5ex}$\Lambda$CDM& 1.341  &0.012&0.008  &0.000011\\
\hline
\end{tabular}}
\label{table6}
\end{table}
\end{center}

\begin{figure}[!h]
\centerline{
  \includegraphics[scale = 0.65]{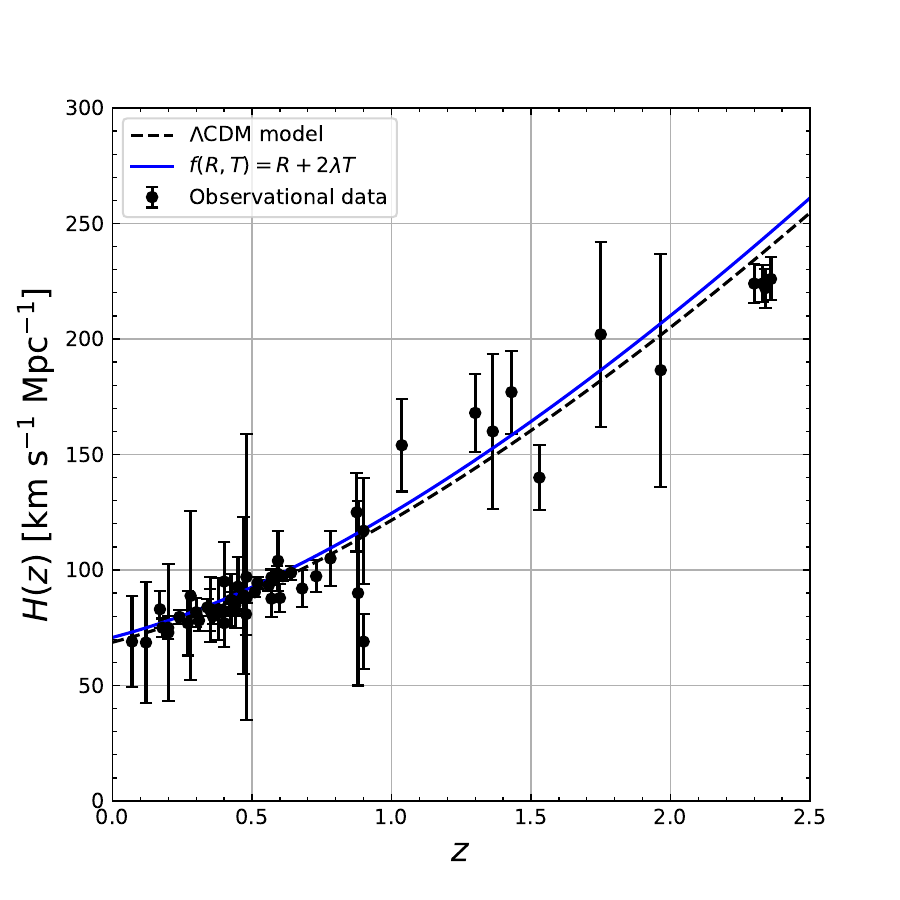}\hspace{0.25cm}
 }
\vspace{-0.2cm}
\caption{Hubble parameter $H(z)$ against cosmological redshift $z$ for the 
constrained set of model parameters for both the $\Lambda$CDM model and 
anisotropic $f(R,T)$-II BIII model along with the observational 
data.}
\label{fig9}
\end{figure}

Moreover, similar to the previous case we have tried to compare the Hubble 
parameter versus cosmological redshift variations for both the $\Lambda$CDM 
and the considered $f(R,T)$-II BIII models with the parameters constrained 
under the combination of $H(z)$ + Pantheon plus + BAO + CMB data set listed in 
Table \ref{table5} as shown in Fig.~\ref{fig9}. The plot shows that for the 
estimated values of the model parameters, the Hubble parameter is consistent
with the observational data. However, the anisotropic BIII Universe in the
considered $f(R,T)$ model shows deviations from the standard $\Lambda$CDM 
Universe which increases with an increase in cosmological redshift $z$. From 
this Fig.~\ref{fig9}, it is also seen that the expansion rate of the 
anisotropic Universe for the considered $f(R,T)$-II BIII model is higher in 
comparison to the standard $\Lambda$CDM model as the redshift value $z$
increases, which is similar to the previous case.

\begin{figure}[!h]
\centerline{
  \includegraphics[scale = 0.6]{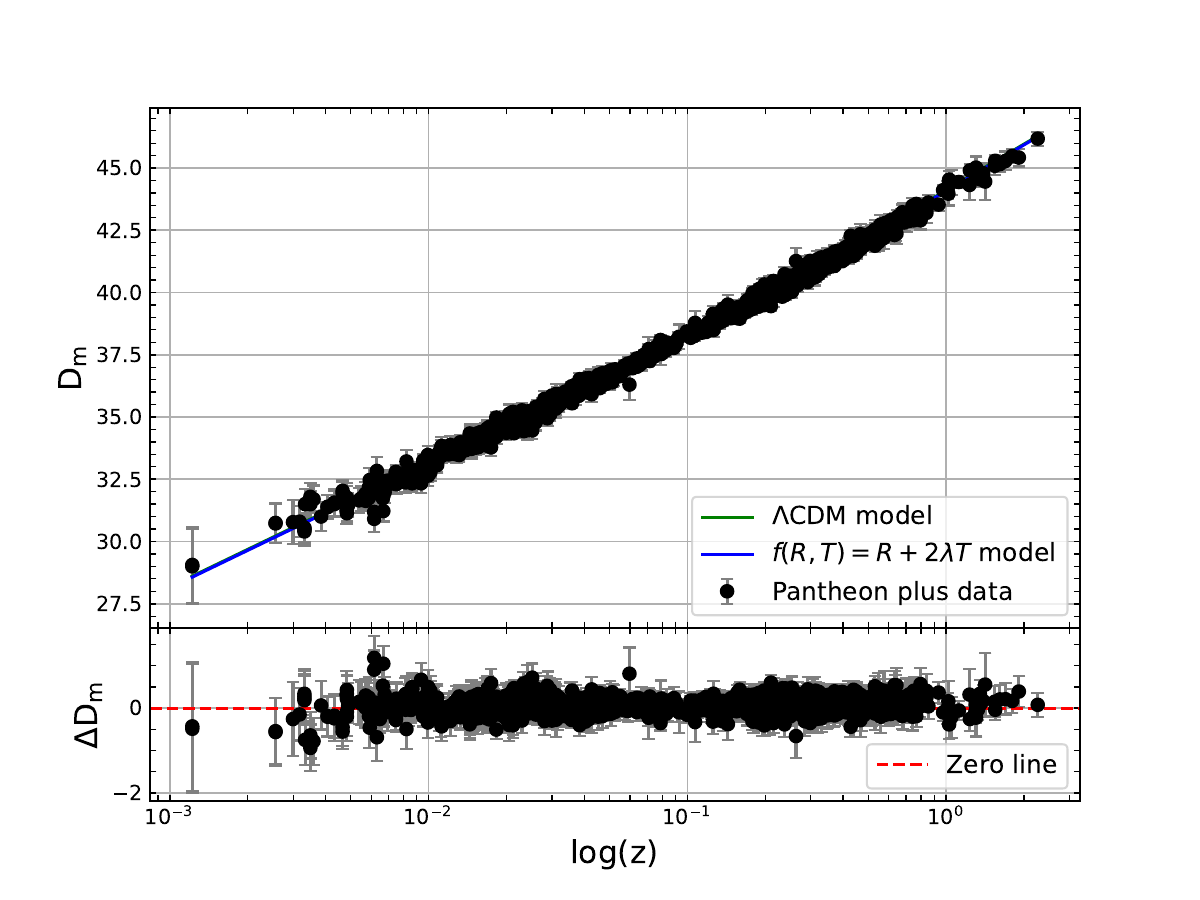}\hspace{0.25cm}
 }
\vspace{-0.2cm}
\caption{Top panel: The Pantheon plus ``Hubble diagram" showing the distance 
modulus $D_m$ versus log of cosmological redshift $z$ for the anisotropic 
$f(R,T)$-II BIII model in comparison with the 
$\Lambda$CDM. Bottom panel: Distance modulus residues against the log of 
cosmological redshift for Pantheon plus data relative to the considered 
anisotropic $f(R,T)$-II BIII model.}
\label{fig10}
\end{figure}

Similarly, we have plotted the distance modulus $D_m$ against cosmological 
redshift $z$ in Fig.~\ref{fig10} for both the $\Lambda$CDM model and the 
anisotropic $f(R,T)$-II BIII model along with distance modulus 
residues relative to $f(R,T)$-II BIII Universe in the logarithmic $z$ scale 
for the constrained set of model parameters as mentioned above in the $H(z)$ 
vs $z$ plot. The plot shows that like $\Lambda$CDM results, the distance 
modulus for the anisotropic Universe for the considered $f(R,T)$-II BIII model 
is consistent with the observational Pantheon plus data obtained from different 
SN Ia for the constrained set of model parameters of Table 
\ref{table5}. Further, the plot of the distance modulus residues also shows 
that the model is consistent with observational data.

\begin{figure}[!h]
\centerline{
  \includegraphics[scale = 0.45]{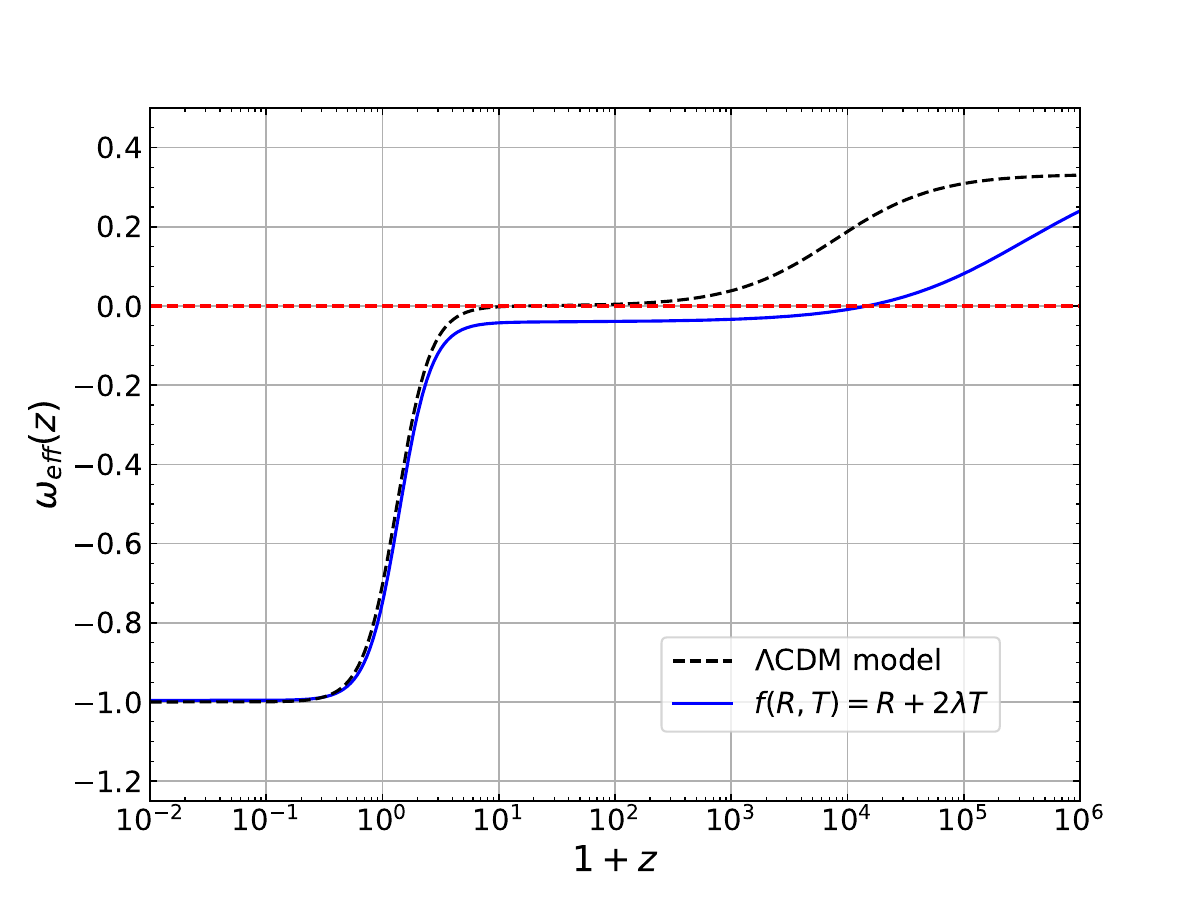}\hspace{0.0cm}
  \includegraphics[scale = 0.45]{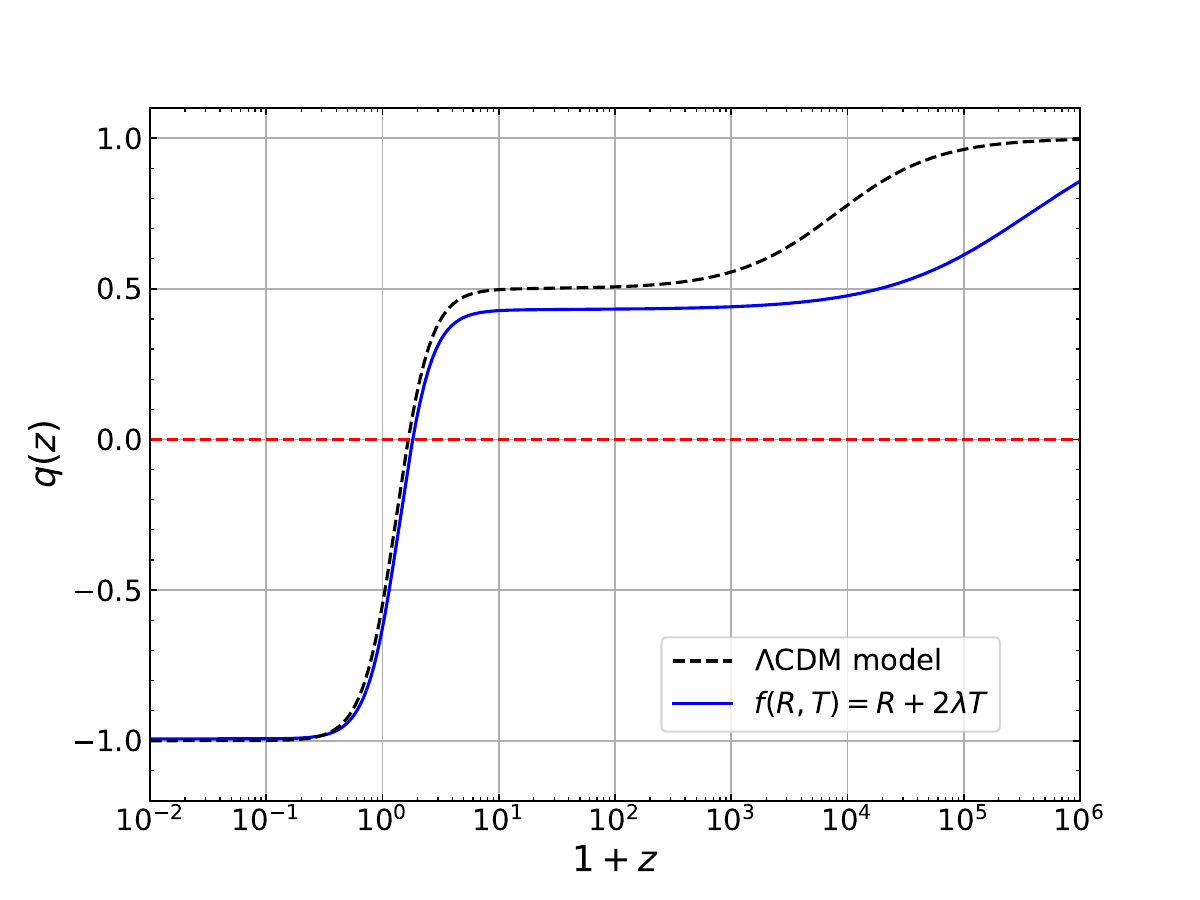}
 }
\vspace{-0.2cm}
\caption{Variation of the effective equation of state $\omega_{eff}$ (left) 
and the deceleration parameter $q$ (right) against cosmological redshift for 
the constrained set of parameters for both the $\Lambda$CDM model and the 
anisotropic $f(R,T)$-II BIII model.}
\label{fig11}
\end{figure}

Apart from these, we have plotted both $\omega_{eff}$ and the deceleration 
parameter $q$ by using equations \eqref{omf_new} and \eqref{dec} against 
cosmological redshift $z$ in Fig.~\ref{fig11} for both the standard 
$\Lambda$CDM model and the anisotropic $f(R,T)$-II BIII model. The anisotropic 
$f(R,T)$-II BIII Universe model plot shows deviations 
from the standard $\Lambda$CDM model results for values of $z>0$ and $z<0$ for 
both $\omega_{eff}$ and $q$ indicating the role of anisotropy in the 
evolution of the Universe as mentioned in the previous case. However, both 
plots show that they are consistent with the standard cosmology at $z = 0$, 
i.e.~in the current time. Thus we can conclude here that the 
$f(R,T) = R + f(T)$ with the considered $f(T) = 2 \lambda T$ is a physically 
viable model to explain the anisotropic Universe.

\subsubsection{$f(R,T) = (\zeta + \eta \tau T)R$}
\begin{figure}[!h]
\centerline{
  \includegraphics[scale = 0.55]{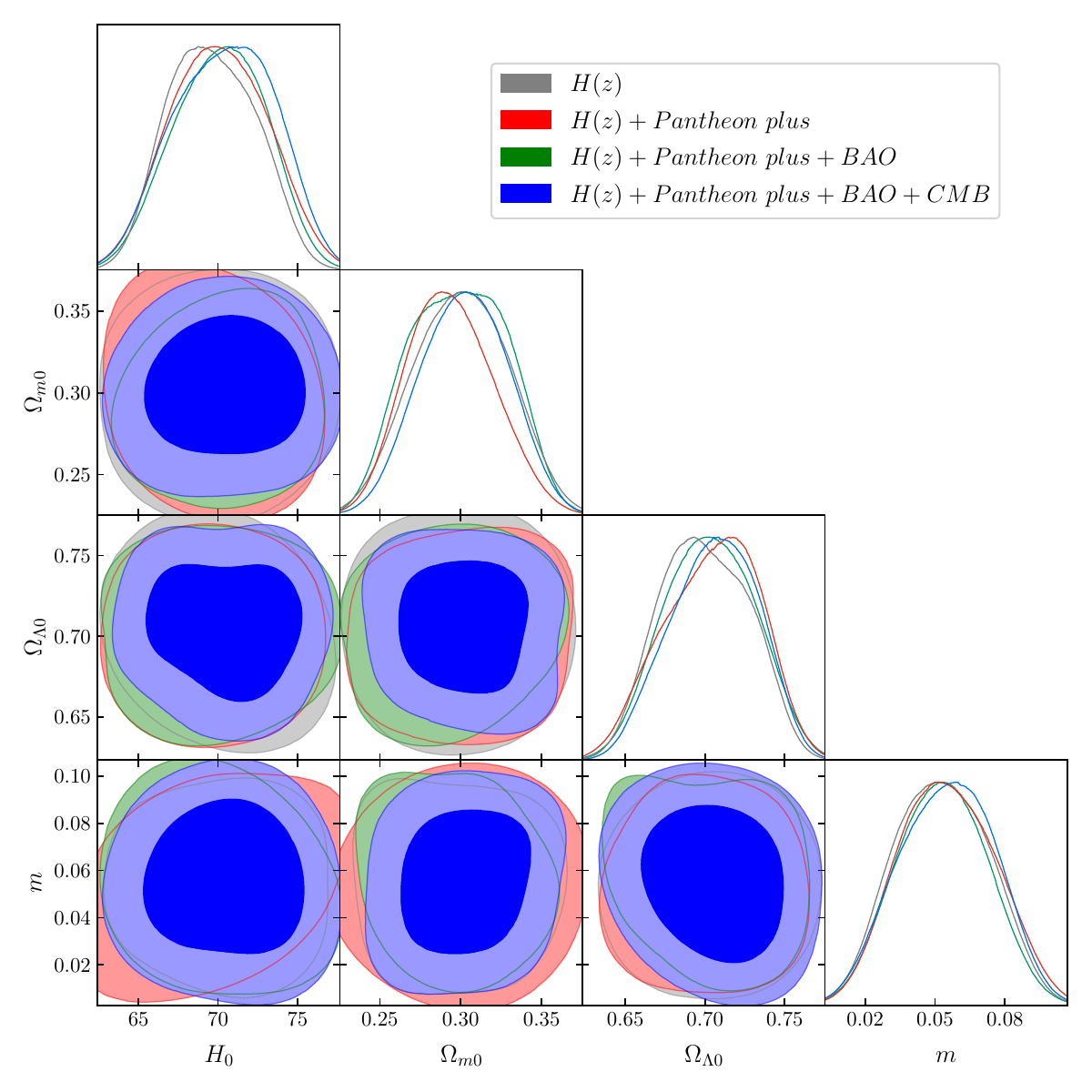}\hspace{0.25cm}
 }
\vspace{-0.2cm}
\caption{One-dimensional and two-dimensional marginalized confidence regions 
($68\%$ and $95\%$ confidence levels) of cosmological and model parameters for 
the $f(R,T)$-III BIII Universe obtained with the help of $H(z)$, Pantheon plus, BAO and CMB data.}
\label{fig12}
\end{figure}

Similar to the previous cases, for convenience, we call this $f(R,T)$ gravity 
model along with the anisotropic BIII metric as the anisotropic $f(R,T)$-III 
BIII model of the Universe. To implement observational constraints on this 
anisotropic $f(R,T)$-III BIII model, again we have taken a multivariate joint 
Gaussian likelihood of equation \eqref{L}. Here also we have considered 
uniform prior distributions for all cosmological parameters and said model 
parameters. The prior ranges of various parameters have been considered as 
follows: $55 < H_0 < 85$, $0.1 <\Omega_{m0} < 0.5$, 
$0.00001 < \Omega_{r0} < 0.0001$, $0.6 < \Omega_{\Lambda 0} < 1$, 
$0.001< m <0.01$, $0.000010<\eta< 0.000015$, $0.0000001 <\tau < 0.0000006$, 
$0.95 <\gamma<1.05$. Here the likelihoods are considered within the mentioned 
ranges such that results should be consistent with the standard Planck data 
release 2018 \cite{}. With these considerations, we have plotted 
one-dimensional and two-dimensional marginalized confidence regions 
($68\%$ and $95\%$ confidence levels) for this anisotropic $f(R,T)$-III BIII 
Universe model and estimate the cosmological parameters along with model 
parameters $H_0$, $\Omega_{mo}$, $\Omega_{\Lambda0}$, $\eta$, $\tau$ 
etc.~within the range of $H(z)$, $H(z)$ + Pantheon plus, $H(z)$ + Pantheon 
plus + BAO and $H(z)$ + Pantheon plus + BAO + CMB data by using the Bayesian 
technique as shown in Fig.~\ref{fig12}. In the same approach, we have 
compiled Table \ref{table7} which shows the constraints ($ 68\%$ confidence 
level) on the considered $f(R,T)$-III BIII model parameters and the 
$\Lambda$CDM model parameters obtained from the different available data sets
by using the Bayesian inference technique. From Table \ref{table7} and 
Fig.~\ref{fig12}, we have found that like in the previous two cases, the 
tightest constraint can be achieved from the joint dataset of $H(z)$ + 
Pantheon~plus + BAO + CMB on all the cosmological parameters for both the 
anisotropic $f(R,T)$-III BIII model and the $\Lambda$CDM model. 

\begin{figure}[!h]
\centerline{
  \includegraphics[scale = 0.5]{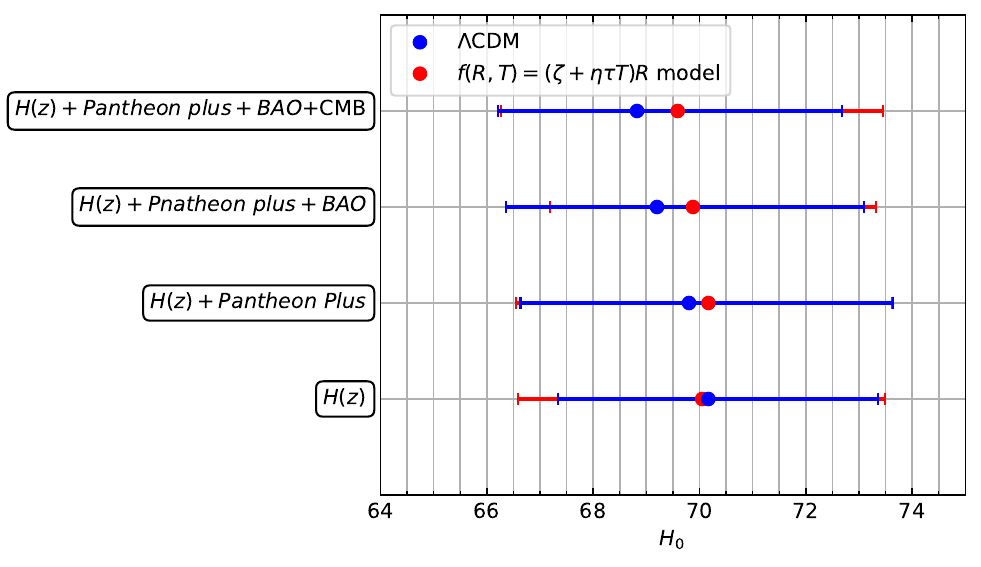}\vspace{0.05cm}
  \includegraphics[scale = 0.5]{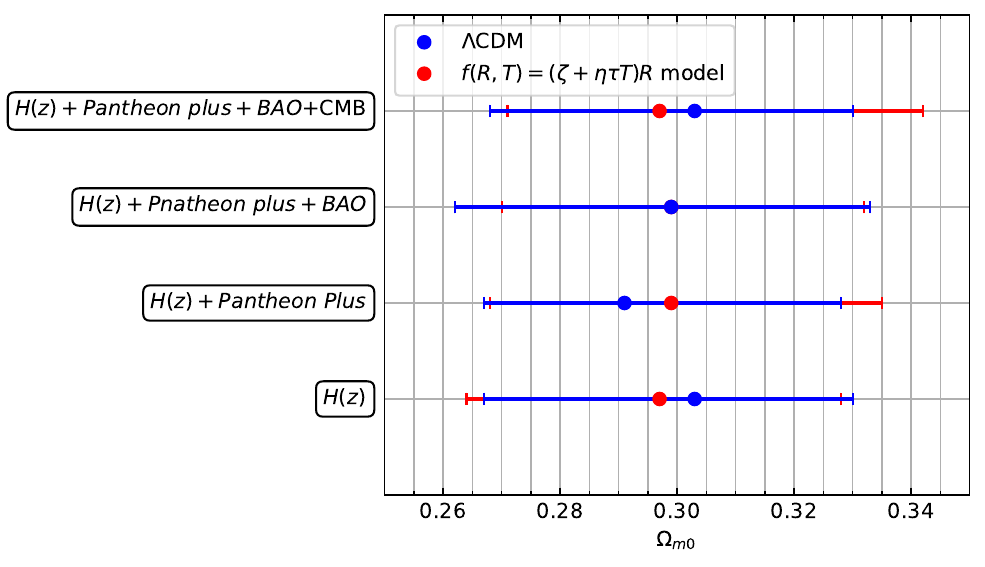}
 }
 \centerline{
  \includegraphics[scale = 0.5]{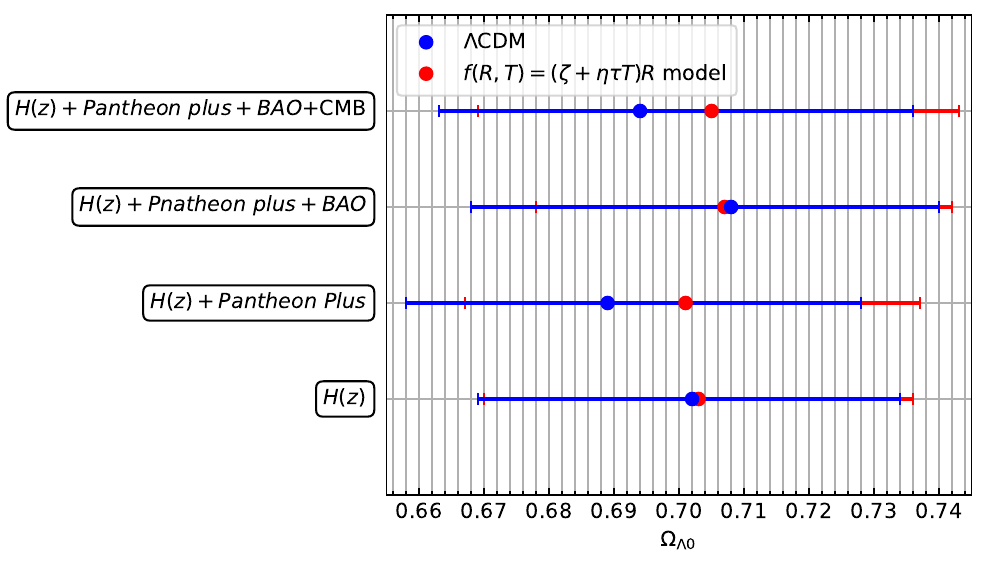}\vspace{0.05cm}
  \includegraphics[scale = 0.5]{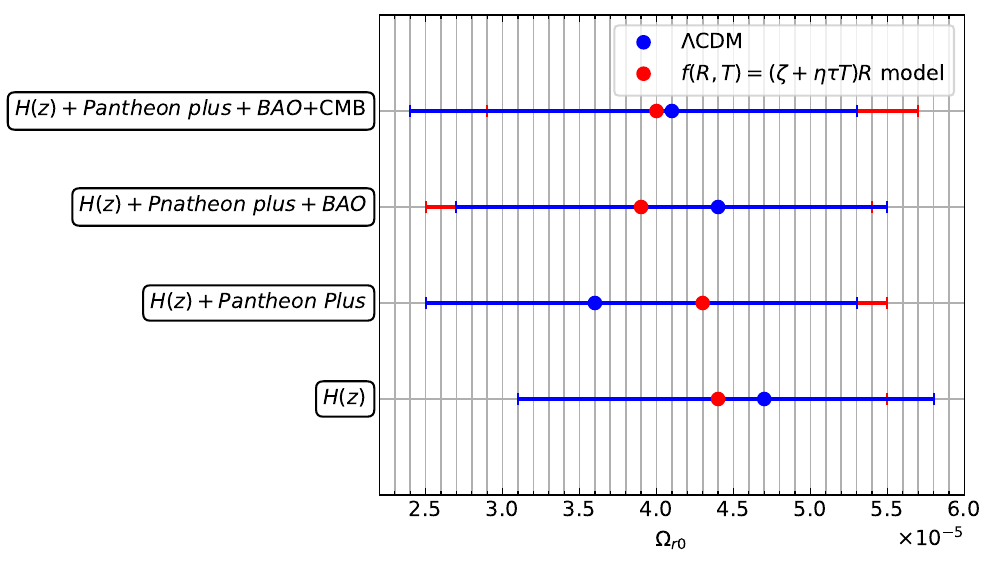}
 }
\vspace{-0.2cm}
\caption{$68\%$ confidence intervals of $H_0$, $\Omega_{m0}$, 
$\Omega_{\Lambda 0}$ and $\Omega_{r0}$ for the anisotropic $f(R,T)$-III BIII 
model in comparison with the $\Lambda$CDM model.}
\label{fig13}
\end{figure}

From Table \ref{table7} we have tried to compare the $H_0$, $\Omega_{m0}
$, $\Omega_{\Lambda0}$ and $\Omega_{r0}$ parameters for both the models for
different data set combinations within $68\%$ confidence intervals as shown
in Fig.~\ref{fig13}. Like in the previous two cases, the shift of the
parameter values from standard $\Lambda$CDM due to the anisotropic background
is clearly observed in the plots of Fig.~\ref{fig13}. The largest deviations of
the cosmological parameters obtianed from these plots are compiled in Table
\ref{table6} for both the standard $\Lambda$CDM model and the considered
anisotropic $f(R,T)$-III BIII cosmological model. From this Table, we again
conclude that the deviations are higher in the $\Lambda$CDM model in comparison
to the anisotropic $f(R,T)$-III BIII model, like in the previous two $f(R,T)$
based models.
\begin{center}
\begin{table}[!h]
\caption{Constrained cosmological parameters including model-specific 
parameters for both anisotropic $f(R,T)$-III BIII model and $\Lambda$CDM 
model obtained through the confidence level corner plots using 
different cosmological data sources.}
\vspace{5pt}
\scalebox{0.88}{
\begin{tabular}{cccccc}
\hline 
\rule[1ex]{0pt}{2.5ex} Model \hspace{2pt} & \hspace{2pt} Parameters \hspace{2pt} & \hspace{2pt} $H(z)$ \hspace{2pt}  & \hspace{2pt} $H(z)$ + Pantheon plus \hspace{2pt} & \hspace{2pt} $H(z)$ + Pantheon plus + BAO \hspace{2pt} & \hspace{2pt} $H(z)$ + Pantheon plus + BAO + CMB \hspace{0.25cm}\\ 
\vspace{-9.5pt}\\
\hline\\[-8pt]
\rule[-2ex]{0pt}{2.5ex} & $H_0$& ${ 70.050^{+3.435}_{-3.457}}$& ${ 70.169^{+3.451}_{-3.619}}$ & ${ 69.878^{+3.435}_{-2.688}}$&${ 69.590^{+3.860}_{-3.326}}$\\
\rule[-2ex]{0pt}{2.5ex} & $\Omega_{m0}$ &${ 0.297^{+0.031}_{-0.033}}$ & ${ 0.299^{+0.036}_{-0.031}}$ &${ 0.299^{+0.033}_{-0.029}}$&${ 0.297^{+0.045}_{-0.026}}$\\
\rule[-2ex]{0pt}{2.5ex} & $\Omega_{r0}$ &${ 0.000044^{+0.000011}_{-0.000013}}$ & ${ 0.000043^{+0.000012}_{-0.000018}}$ &${ 0.000039^{+0.000015}_{-0.000014}}$& ${ 0.000040^{+0.000017}_{-0.000011}}$\\
\rule[-2ex]{0pt}{2.5ex} & $\Omega_{\Lambda0}$&${ 0.703^{+0.033}_{-0.033}}$&${  0.701^{+0.036}_{-0.034}}$&${ 0.707^{+0.035}_{-0.029}}$&${ 0.705^{+0.038}_{-0.036}}$\\
\rule[-2ex]{0pt}{2.5ex} $f(R,T)$-III BIII & $m$&${ 0.045^{+0.011}_{-0.009}}$&${  0.051^{+0.028}_{-0.021}}$ & ${ 0.055^{+0.024}_{-0.027}}$ &${ 0.056^{+0.023}_{-0.031}}$\\
\rule[-2ex]{0pt}{2.5ex} & $\eta$ &${ 0.000015^{+0.0000009}_{-0.0000011}}$ &${  0.000014^{+0.0000012}_{-0.0000011}}$ &${ 0.000013^{+0.0000012}_{-0.0000013}}$&${ 0.000015^{+0.0000017}_{-0.0000015}}$ \\
\rule[-2ex]{0pt}{2.5ex} & $\tau$ &${ 0.00000041^{+0.00000016}_{-0.00000011}}$ & ${ 0.00000039^{+0.00000013}_{-0.00000011}} $ &${ 0.00000038^{+0.00000013}_{-0.00000019}}$&${ 0.00000036^{+0.00000012}_{-0.00000016}}$ \\
\rule[-2ex]{0pt}{2.5ex} & $\gamma$ &${ 0.998^{+0.012}_{-0.017}} $&${ 0.996^{+0.015}_{-0.016} }$ &${ 0.994^{+0.018}_{-0.016}}$&${ 0.995^{+0.017}_{-0.013}}$\\
\rule[-2ex]{0pt}{2.5ex} & $\zeta$ &${ 0.995^{+0.011}_{-0.013}}$ &${ 0.994^{+0.018}_{-0.016}}$&${ 0.997^{+0.011}_{-0.015}}$&${ 0.997^{+0.017}_{-0.013}}$\\
\hline\\[-8pt]
\rule[-2ex]{0pt}{2.5ex} & $H_0$&$70.167^{+3.192}_{-2.823}$ & $69.804^{+3.841}_{-3.169}$&$69.202^{+3.893}_{-2.833}$&$68.826^{+3.857}_{-2.620}$\\
\rule[-2ex]{0pt}{2.5ex} $\Lambda$CDM & $\Omega_{m0}$ &$0.303^{+0.027}_{-0.036}$&$0.291^{+0.037}_{-0.024}$ &$0.299^{+0.034}_{-0.037}$&$0.303^{+0.027}_{-0.035}$\\
\rule[-2ex]{0pt}{2.5ex} & $\Omega_{r0}$ &$0.000047^{+0.000011}_{-0.000016}$ & $0.000036^{+0.000017}_{-0.000011}$ &$0.000044^{+0.000011}_{-0.000017}$&$0.000041^{+0.000012}_{-0.000017}$\\
\rule[-2ex]{0pt}{2.5ex} &  $\Omega_{\Lambda0}$&$0.702^{+0.032}_{-0.033}$& $0.689^{+0.039}_{-0.031}$ &$0.708^{+0.032}_{-0.040}$&$0.694^{+0.042}_{-0.031}$\\
\hline
\end{tabular}}
\label{table7}
\end{table}
\end{center}
\begin{center}
\begin{table}[!h]
\caption{Deviations of cosmological parameters for different combinations of 
data sets for the $\Lambda$CDM and the anisotropic $f(R,T)$-III BIII model.}
\vspace{8pt}
\scalebox{1}{
\begin{tabular}{ccccc}
\hline 
\rule[0.1ex]{0pt}{2.5ex}Model \hspace{0.25cm} &\hspace{0.25cm} $\Delta H_0$ \hspace{0.25cm}  & \hspace{0.25cm} $\Delta \Omega_{m0} $ \hspace{0.25cm} & \hspace{0.25cm} $\Delta \Omega_{\Lambda0}$ \hspace{0.25cm} &\hspace{0.25cm} $\Delta \Omega_{r0}$ \hspace{0.25cm}\\ 
\hline
\rule[0.1ex]{0pt}{2.5ex}$f(R,T)$-III BIII& {0.579} &{0.002} &{0.004} &{0.000005}\\
\rule[0.1ex]{0pt}{2.5ex}$\Lambda$CDM& 1.341  &0.012&0.008  &0.000011\\
\hline
\end{tabular}}
\label{table8}
\end{table}
\end{center}
\begin{figure}[!h]
\centerline{
  \includegraphics[scale = 0.65]{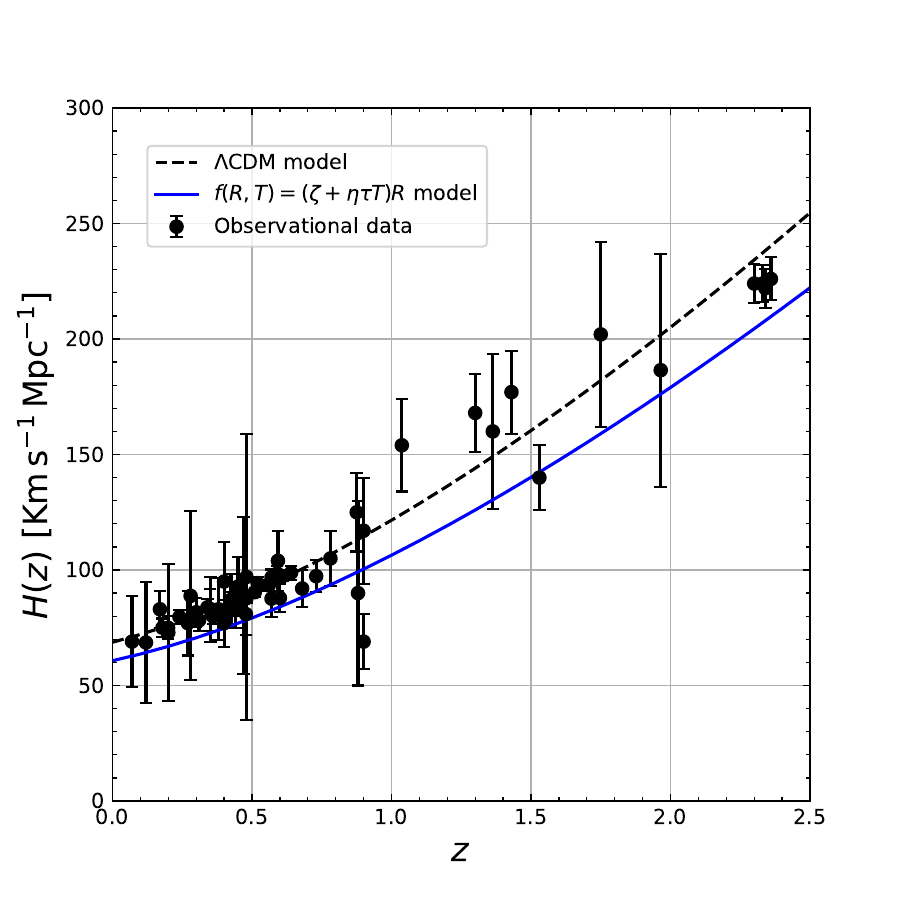}\hspace{0.25cm}
 }
\vspace{-0.2cm}
\caption{Hubble parameter $H(z)$ against cosmological redshift $z$ for the 
constrained set of model parameters for both the $\Lambda$CDM and the 
anisotropic $f(R,T)$-III BIII model along with the observational data.}
\label{fig14}
\end{figure}

As in the previous cases, we have tried to compare the Hubble parameter versus 
cosmological redshift variations for both the $\Lambda$CDM and the considered 
$f(R,T)$-III BIII models using the parameters constrained within the 
combination of $H(z)$ + Pantheon plus + BAO + CMB data listed in Table 
\ref{table7} as shown in Fig.~\ref{fig14}. This figure shows that for the 
estimated values of cosmological parameters, the Hubble parameter is 
consistent with the observational data. However, the anisotropic $f(R,T)$-III 
BIII model shows deviations from the standard $\Lambda$CDM plot with the 
increase of cosmological redshift $z$. From Fig.~\ref{fig14}, we have found 
that the expansion rate of the anisotropic $f(R,T)$-III BIII Universe is
lower in comparison to the standard $\Lambda$CDM model as the redshift value 
$z$ increases, which differs from the previous two models in which the 
anisotropic Hubble expansion was higher.

\begin{figure}[!h]
\centerline{
  \includegraphics[scale = 0.6]{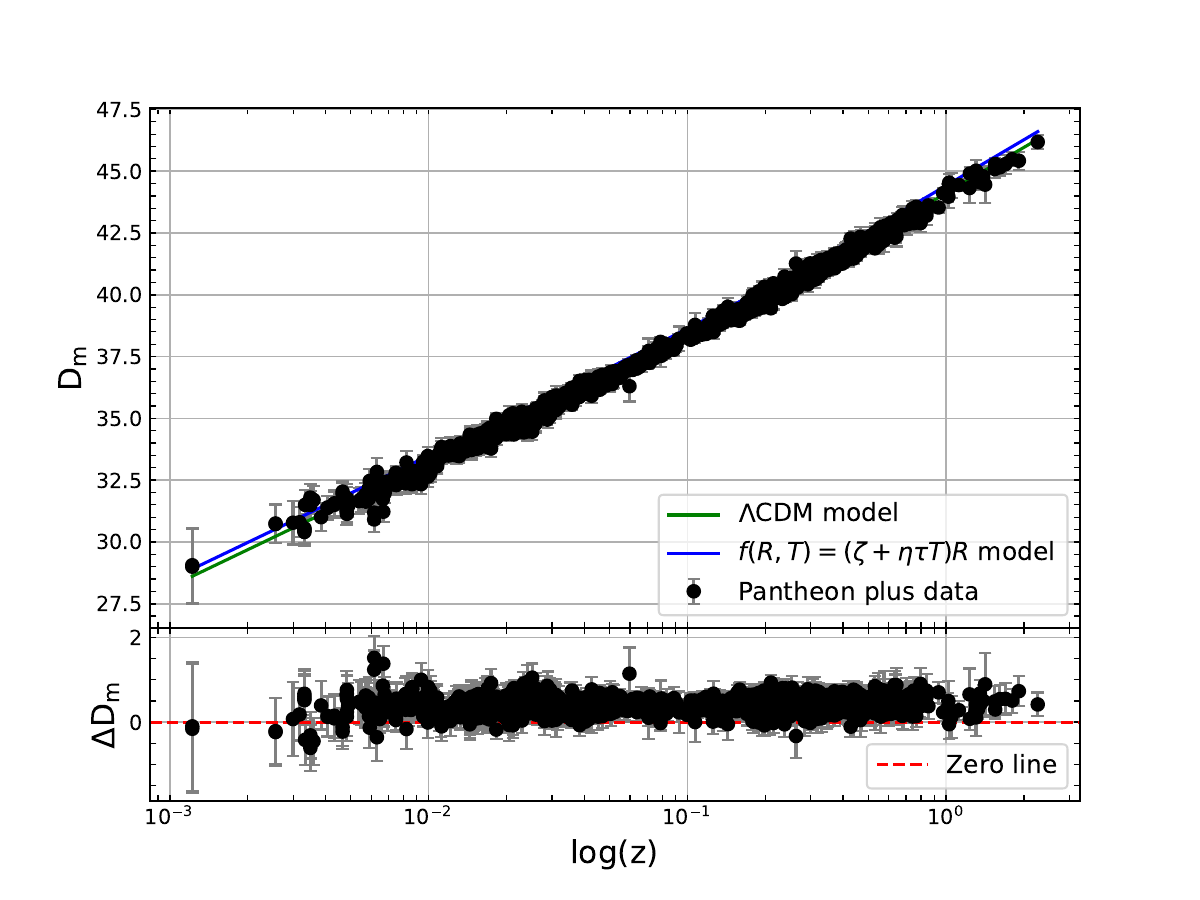}\hspace{0.25cm}
 }
\vspace{-0.2cm}
\caption{Top panel: The Pantheon plus ``Hubble diagram" showing the distance 
modulus $D_m$ versus log of cosmological redshift $z$ along with the 
$\Lambda$CDM model and the anisotropic $f(R,T)$-III BIII model results. Bottom 
panel: Distance modulus residues against the log of cosmological redshift for 
Pantheon plus data relative to anisotropic $f(R,T)$-III BIII model.}
\label{fig15}

\end{figure}

Along with the Hubble parameter, we have also plotted the distance modulus 
$D_m$ against the log of cosmological redshift $z$ in Fig.~\ref{fig15} for 
both the 
$\Lambda$CDM model and anisotropic $f(R,T)$-III BIII model along with the 
distance modulus residues relative to the $f(R,T)$-III BIII Universe in the 
logarithmic $z$ scale for the constrained set of model parameters as did in 
the previous two models. The plot shows that like $\Lambda$CDM results, the 
distance modulus for the anisotropic $f(R,T)$-III BIII Universe is consistent 
with the observational Pantheon plus data obtained from different Type Ia 
supernovae (SN Ia) for the constrained set of model parameters of Table 
\ref{table7}. Further, the plot of the distance modulus residues also shows 
that the model is consistent with observational data.
\begin{figure}[!h]
\centerline{
  \includegraphics[scale = 0.45]{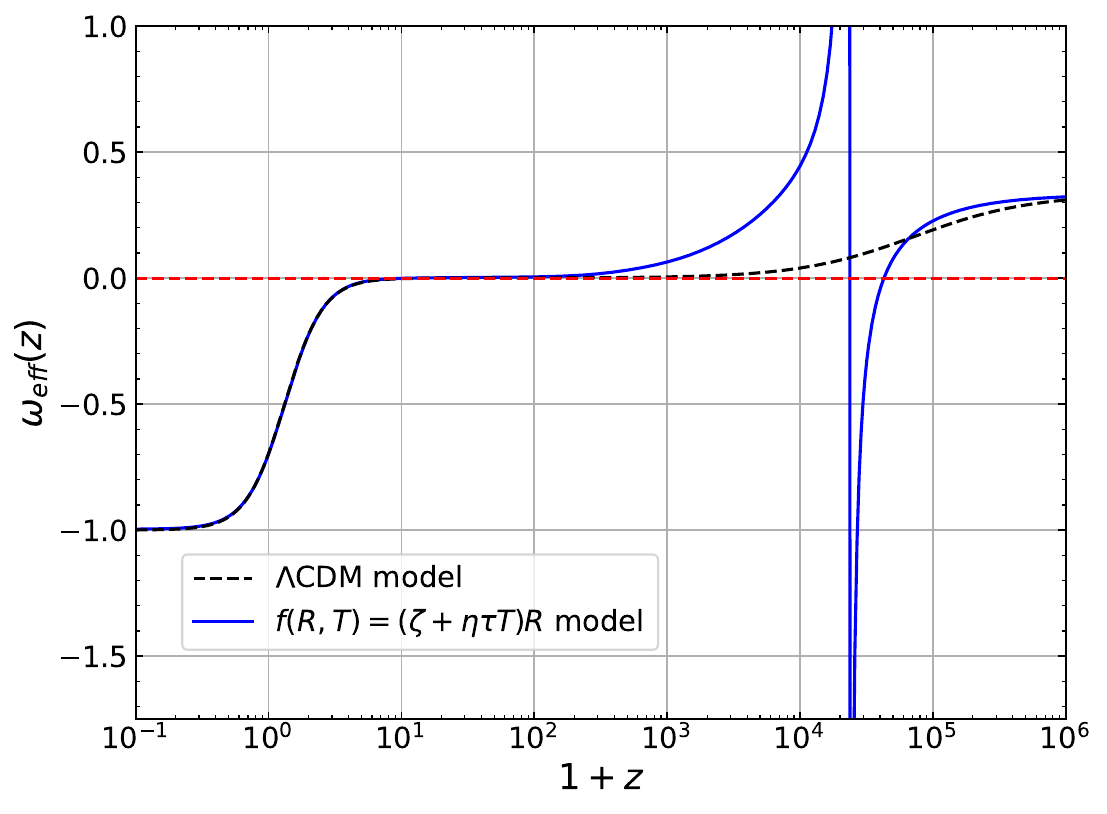}\hspace{0.0cm}
  \includegraphics[scale = 0.45]{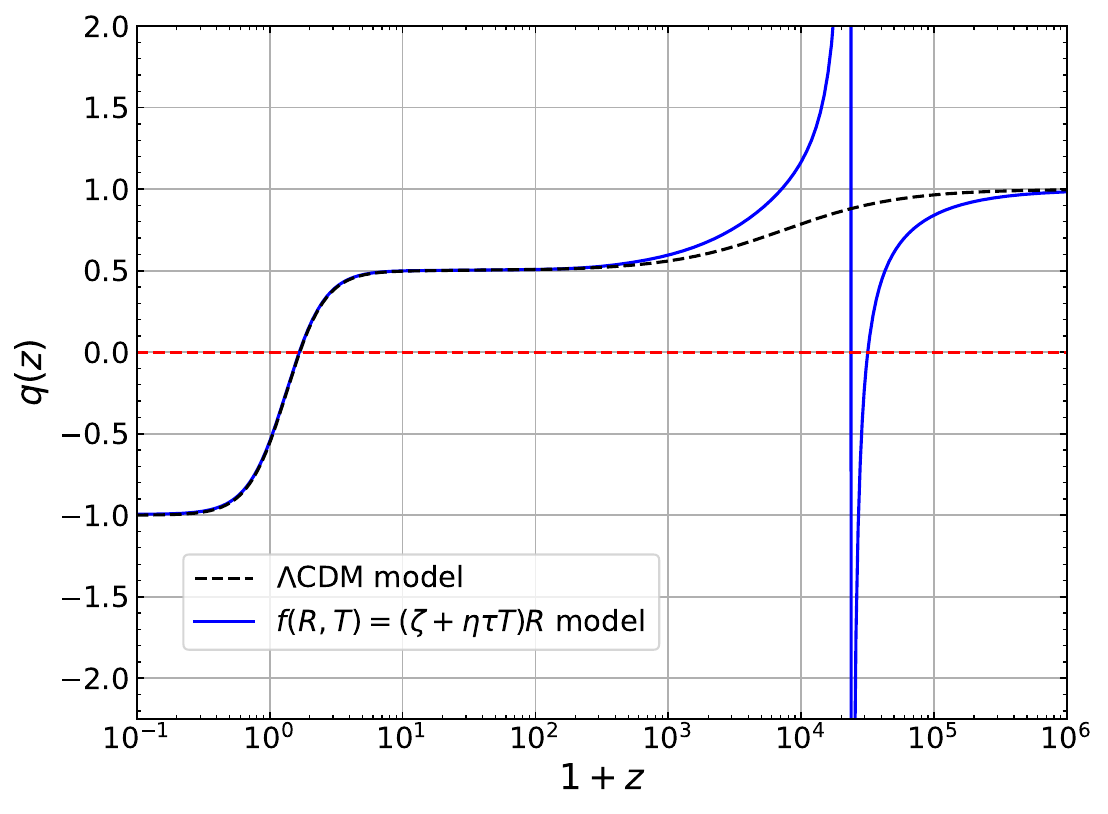}
 }
\vspace{-0.2cm}
\caption{Variation of the effective equation of state $\omega_{eff}$ (left) and 
deceleration parameter $q$ (right) against cosmological redshift for the 
constrained set of model parameters for both $\Lambda$CDM and anisotropic
$f(R,T)$-III BIII models.}
\label{fig16}
\end{figure}

Furthermore, we have plotted the effective equation of state $\omega_{eff}$ from 
the equation \eqref{Om_eff3} against cosmological redshift $z$ for the 
constrained set of model parameters listed in Table \ref{table7} for both 
the $\Lambda$CDM model and the anisotropic $f(R,T)$-III BIII model in Fig 
\ref{fig16} (left). The plot shows a sharp discontinuity in the 
{radiation-dominated} period leading to the sharp deviations from the standard 
$\Lambda$CDM results and hence the considered $f(R,T) = (\zeta + \eta \tau T)R$
model is not suitable for studying the evolution of the Universe. 
Similarly the deceleration parameter $q$ of equation \eqref{dec} against 
cosmological redshift $z$ plot in Fig.~\ref{fig16} (right) for anisotropic 
$f(R,T)$-III BIII model shows a sharp discontinuity in the 
{radiation-dominated} region and hence a strong deviation from standard 
$\Lambda$CDM results has been observed. Thus the model is not viable for 
studying the evolution of the Universe especially the 
{radiation-dominated} region even though the model shows consistent 
results with the observable data at $z = 0$ or near past.
    
\section{Summary and Conclusions}\label{5}
In this work, we have considered the BIII metric in the $f(R,T)$ 
gravity theory and are trying to understand its effect on the cosmological 
parameters and hence the evolutions of the Universe by using three different 
$f(R,T)$ gravity models. We have started our work by considering the general 
form of field equations for the $f(R,T)$ theory of gravity and all the related 
equations and expressions in Section \ref{2}, which are required us to carry 
forward our work. In Section \ref{3} we have derived the field equations in 
the $f(R,T)$ gravity for the BIII metric.  Here we have considered three 
$f(R,T)$ models and derive field equations for each model for the conventional 
energy-momentum tensor $T_{\mu \nu}$. From these field equations, we have 
further derived various cosmological parameters for each of the $f(R,T)$ models.
   
In Section \ref{4} we have discussed the method of Bayesian inference used to
constrain the cosmological parameters. We started the section with the
general formulation of Bayesian inference and then we discussed the various 
observational data compilations, viz., Hubble parameter $H(z)$ data, SN Ia 
data, BAO data and CMB data, and their constraining techniques 
using the Bayesian method along with their respective likelihoods in several
subsections. Further, we have constrained the various cosmological parameters
like $H_0$, $\Omega_{m0}$, $\Omega_{r0}$, $\Omega_{\Lambda0}$ along with 
the parameter $m$ which is the BIII metric parameter and other model 
parameters by using mentioned the Bayesian inference technique within those 
observational data for all the three $f(R,T)$ models and the estimated values 
of these constrained parameters are listed in three tables, Table \ref{table3},
Table \ref{table5} and Table \ref{table7} along with the corresponding
values for the standard $\Lambda$CDM model. We have estimated these values by 
using one-dimensional and two-dimensional marginalized $68\%$ and $95\%$ 
confidence level corner plots obtained by employing the Bayesian technique.

With the constrained set of current values of cosmological parameters and 
model parameters, we have plotted the Hubble parameter and distance modulus 
along with the available observational data. For the Hubble parameter, we 
have used the expression of \eqref{hub_new1}, \eqref{hub_new2} and 
\eqref{hub_new3} obtained from the three $f(R,T)$ models along with the 
$\Lambda$CDM model's results. We have found that for all three models, 
the Hubble parameter plots are consistent with the observational data. 
However, we have observed that for $f(R,T) = \alpha R + \beta \lambda T$ 
and $f(R,T) = R + 2f(T)$ models, i.e.~for anisotropic $f(R,T)$-I BIII model and
$f(R,T)$-II BIII model respectively, the Hubble expansion rate is higher than 
the standard $\Lambda$CDM model's results for all values of cosmological 
redshift $z$, whereas for $f(R,T) = (\zeta + \eta \tau T)R$ model, i.e.~for 
anisotropic $f(R,T)$-III BIII model the Hubble expansion rate is lower than 
the standard $\Lambda$CDM results for values of $z>0$.
  
Apart from the Hubble parameter, we have plotted the distance modulus for the 
constrained values of cosmological and model parameters for each three 
models against the log of $z$ with Pantheon plus data and the $\Lambda$CDM 
model's results. We have also plotted the residues of the distance modulus 
with respect to the log of $z$. These plots show good agreement with the 
standard cosmology and are 
consistent with observational data. The goodness of fitting can also be 
observed in distance modulus residue plots for all three models. We have also 
plotted the effective equation of state $\omega_{eff}$ and deceleration 
parameter $q$ against cosmological redshift $z$ for all the three $f(R,T)$
BIII models along with that for the $\Lambda$CDM model. For the first two 
models i.e.~for $f(R,T)$-I BIII and $f(R,T)$-II BIII models, both 
$\omega_{eff}$ and $q$ are consistent with the $\Lambda$CDM results at $z=0$, 
however deviations from the standard $\Lambda$CDM result has been observed 
for the values of $z>0$ which indicate the effects of anisotropic background 
during the evolutions of different phases of the Universe. On the other hand, 
there is an interesting result observed in $\omega_{eff}$ against cosmological 
redshift $z$ plot for the $f(R,T)$-III BIII model. The plot shows that there 
is a sudden blown-up of $\omega_{eff}$ at the {radiation-dominated} 
region and this discontinuity raised the question of the viability of the 
$f(R,T) = (\zeta + \eta \tau T)R$ model. Similarly, in $q$ versus $z$ plot also
we have observed the same discontinuity and thus we may arrive at a conclusion 
that the $f(R,T)= f_1 (R)+ f_2(R)f_3 (T)$ with $f_1(R) = \zeta R$, 
$f_2(R) = \eta R$ and $f_3 (T) = \tau T$ is not a physically viable model to 
study cosmological evolution of the BIII Universe.
  
Finally, we have observed some important results in the study of the BIII 
Universe in 
$f(R,T)$ gravity and constrained several cosmological parameters through 
using observational data by employing the Bayesian inference technique. Here 
we have used three $f(R,T)$ models and to avoid complexity we have considered 
the functions of $f(R)$ and $f(T)$ in linear forms. In the first two models 
i.e.~$f(R,T) = \alpha R + \beta f(T)$ and $f(R,T) = R + 2 \lambda f(T)$, the 
expressions are relatively simple and thus the cosmological parameters are 
easy to calculate. However, in the third model, the Ricci scalar $R$ and 
trace of the energy-momentum tensor $T$ are appeared in the resulting 
cosmological parameters' expressions and hence are not easy to deal with. In 
our study, we 
have expressed both $R$ and $T$ in terms of cosmological redshift $z$ and 
density parameters which help us to carry forward our analysis. But the 
complexity still persists. Again we have considered the 
$\sigma^2 \propto \theta^2$ assumptions and hence avoided  anisotropic density 
parameters in  the expressions as the observational data do not have
much information in these regards. But cosmological parameters like $H_0$, 
$D_m$, $\omega_{eff}$, $q$ etc.~have shown deviations from the standard 
results and thus a clear signature of the anisotropic metric background has 
been indicated. Further, the third $f(R,T)$ model we have considered is found to
be inconsistent with standard cosmology at the {radiation-dominated} 
phase and we can discard this type of model in our future study of the Bianchi 
Universe while studying its 
physical viability. The other two models suggest the presence of anisotropy in 
the early Universe as the various model parameters shifted from the standard 
$\Lambda$CDM results. To confirm its presence in a more concrete way we need 
more observational data of the early Universe. In this regard, the Thirty 
Meter Telescope \cite{tmt}, Extremely Large Telescope \cite{elt}, Cherenkov 
Telescope Array \cite{cta} and other similar projects may help physicists to 
understand the early-stage scenario of the Universe in the near future.

\appendix*
\section{Derivation of the Ricci Scalar expression for the $\boldsymbol{f(R,T) = (\zeta + \eta \tau T)R}$ model}

{For the BIII metric \eqref{metric}, the general expression of the Ricci 
scalar can be written as}
\begin{equation}\label{ricci}
{R = 2\left[\frac{\dot{a_1}}{a_1} \frac{\dot{a_2}}{a_2} +\frac{\dot{a_2}}{a_2} \frac{\dot{a_3}}{a_3} + \frac{\dot{a_3}}{a_3} \frac{\dot{a_1}}{a_1} + \frac{\ddot{a_1}}{a_1}+ \frac{\ddot{a_2}}{a_2}+ \frac{\ddot{a_3}}{a_3} - \frac{m^2}{a_1^2}\right].}
\end{equation}
{We can re-rewrite equation \eqref{ricci} in terms of the directional 
Hubble parameters with the condition $H_1 = H_2$, which takes the form:}
\begin{equation}
{R = 2\left[3 H_1^2+2H_1H_3 + H_3^2+ (2\dot{H_1}+\dot{H_3}) - \frac{m^2}{a_1^2}\right]}.
\end{equation}
{By rearranging the above equation, we can write}
\begin{equation}
{R = \{H_1^2+2H_1H_3 -\frac{m^2}{a_1^2}\}+ 2\{H_1^2+ H_3^2 + H_1 H_3 + (\dot{H_1} + \dot{H_3})\} + \{3H_1^2+ 2\dot{H_1} - \frac{m^2}{a_1^2}\}.}
\end{equation}
{Now, using equations \eqref{Fe1}, \eqref{Fe2} and \eqref{Fe3} we can 
write the Ricci scalar as}
\begin{equation}\label{r1}
{R = \frac{(\rho-3p)+ \eta \tau R (\rho - 7 p)}{(\zeta+\eta \tau T)}
 = \frac{\rho -3p}{\{\zeta +2\eta \tau (5p-\rho)\}}}.
\end{equation}
{Here, we have used $T = -\rho + 3p$. Furthermore, we can use the 
relation $p=\omega\rho$ in equation \eqref{r1} and hence we can write}
\begin{equation}
{R = \frac{(1-3\omega)\rho}{\{\zeta + 2 \eta \tau (5\omega -1)\rho\}}}.
\end{equation}
{Finally, using the equation \eqref{solr}, \eqref{soll} and 
\eqref{solmn} for $\omega$ equal to $1/3$, $-1$ and $0$ respectively, and 
$a = \frac{1}{(1+z)}$, we can obtain the expression of Ricci scalar as}
\begin{equation}\label{R}
{R(z) = \frac{3H_0^2\left\{\zeta \Omega_{m0}(1+z)^3 + 4\,\Omega_{\Lambda0}\right\}}{\zeta + 6\eta\, \tau H_0^2\left\{-\zeta \Omega_{m0}(1+z)^3 +\frac{2}{3}\,\Omega_{r0}(1+z)^4 -6\,\Omega_{\Lambda 0}\right\}}.}
\end{equation} 
\section*{ACKNOWLEDGEMENT}
UDG is thankful to the Inter-University Centre for Astronomy and Astrophysics 
(IUCAA), Pune, India for the Visiting Associateship of the institute.

\bibliographystyle{apsrev}

\end{document}